   \documentclass[preprint]{aastex}

\newcommand{\etal}{\it{et al.}\rm}

\slugcomment{To appear in the Astronomical Journal, August 2002}
\shorttitle{The Post-Red Supergiant IRC+10420}
\shortauthors{Humphreys et al.}

\begin{document}

\title{Crossing the `Yellow Void' -- Spatially Resolved Spectroscopy of the Post--Red 
Supergiant IRC+10420 and Its Circumstellar Ejecta\altaffilmark{1}}

\author{Roberta M. Humphreys\altaffilmark{2}, Kris Davidson, and Nathan Smith\altaffilmark{2}}

\affil{Astronomy Department, University of Minnesota, Minneapolis, MN 55455}

\email{roberta@aps.umn.edu}

\altaffiltext{1}{Based on observations made with the NASA/ESA {\it
Hubble Space Telescope}, obtained at the Space Telescope Science
Institute, which is operated by the Association of Universities for
Research in Astronomy, Inc., under NASA contract NAS5-26555.}                         

\altaffiltext{2}{Visiting Astronomer, Kitt Peak National Observatory (KPNO), National 
Optical Astronomy Observatory , which is operated by the Association of Universities 
for Research in Astronomy, Inc., under cooperative agreement with the National Science 
Foundation.}

\begin{abstract}
IRC +10420 is one of the extreme hypergiant stars  that define the 
empirical upper luminosity boundary in the HR diagram. During their
post--RSG evolution, these massive stars enter a temperature range 
(6000-9000$\arcdeg$K) of increased dynamical instability, high mass 
loss, and increasing opacity, a  semi--forbidden region, that de Jager 
and his collaborators have called the ``yellow void''.  We report HST/STIS 
spatially resolved spectroscopy of IRC +10420 and its reflection nebula 
with some surprising results. Long slit spectroscopy of the reflected 
spectrum allows us to effectively view the star from different 
directions.  Measurements of the double--peaked H$\alpha$ emission profile 
show a uniform outflow of gas in a nearly spherical distribution, contrary 
to previous models with an equatorial disk or bipolar outflow.  Based 
on the temperature and mass loss rate estimates that are usually quoted 
for this object, the wind is optically thick to the continuum at some 
and possibly all wavelengths.  Consequently the observed variations 
in apparent spectral type and inferred temperature 
are changes in the wind and do not necessarily 
mean that the underlying stellar radius and interior 
structure are evolving on such a short timescale.  To explain the evidence 
for simultaneous outflow and infall of material near the star, we propose 
a ``rain'' model in which blobs of gas condense in regions of lowered 
opacity outside the dense wind.  With the apparent warming of its wind, 
the recent appearance of strong emission, and a decline in the mass loss 
rate, IRC +10420 may be about to shed its opaque wind, cross the 
``yellow void'', and emerge as a hotter star. 
\end{abstract}

\keywords{stars:atmospheres --- stars:evolution --- stars:individual(IRC+10420) --- stars:supergiants }

\section{Introduction}

  IRC +10420 may be one of the most important stars in the HR diagram 
for understanding  the final stages of massive star evolution.
With its high luminosity of $L \ {\sim} \ 5 {\times} 10^6$ $L_{\odot}$
(Jones et al.\ 1993, Paper I) and extraordinary mass loss rate of 
3 -- 6 $\times$ 10$^{-4}$ $M_{\odot}$ yr$^{-1}$
\citep{Knapp85,Oud96,RMH97},  IRC+10420 is one of the few known 
stars that define the empirical upper luminosity boundary in the HR diagram
at intermediate temperatures, between the main sequence and the red supergiant
region (see Figure 9 in \citet{HD94} and Figure 14 in this paper).  
The evolutionary state of these cool hypergiants is not 
obvious;  they may be evolving toward the red supergiant region 
or  back to the blue side after having 
lost considerable mass as  red supergiants.   As one of 
the warmest known OH masers and  one of the brightest 10--20 $\mu$m 
IR sources in the sky, together with its A-F--type spectrum and visible
ejecta \citep[Paper II]{RMH97}, IRC +10420 
is an excellent candidate for post-red supergiant evolution.

  It has been variously described in the literature as either a true supergiant 
\citep{RMH73,Gig76,Mut79} 
or  a proto-planetary/post-AGB star \citep{Hab89,Hri89,Bow89},
depending on distance estimates that ranged from 1.5 to 7 kpc. 
In Paper I we combined multi-wavelength spectroscopy, photometry, 
and polarimetry to confirm a large distance of 4--6 kpc and the 
resulting high luminosity mentioned above.  We showed that its large 
optical--IR polarization and color excess are mostly interstellar, 
not circumstellar. These  conclusions were later supported by 
\citet{Oud96} who demonstrated from CO data that IRC+10420 has
to be much more luminous than  the AGB limit. In Paper I
we concluded that IRC +10420 is a post--red supergiant evolving 
back toward the blue side of the HR diagram, in an evolutionary phase 
analogous to the proto-planetary/post-AGB stage for lower mass stars.    
  \citet{deJ98} has suggested that most
if not all of the intermediate temperature hypergiants are post--red
supergiants. In their blueward evolution these
very massive stars enter a temperature range (6000--9000$\arcdeg$K) of
increased dynamical instability, a semi-forbidden
region in the HR diagram, that he called the {\it ``yellow void''}, where high 
mass loss episodes occur. 

  Our HST images of IRC+10420 (Paper II) revealed a 
complex circumstellar environment, with a variety of  structures including
condensations or knots, ray-like features, and semi--circular arcs or loops 
within  2$\arcsec$ of the star, plus one or more distant reflection shells. 
A few other intermediate-temperature hypergiants such as $\rho$ Cas 
and HR 8752 occupy the same region in the HR diagram, 
but IRC +10420 is the only one with obvious circumstellar nebulosity, 
making it our best candidate for  a star in transition  from a red supergiant
possibly to an  S Dor--type  variable (LBV), a Wolf-Rayet star, or a pre-SN 
state.   Moreover, its photometric history (\citet{Got78}, and our Paper I) 
and spectroscopic variations (Paper I and \citet{Oud96,Oud98}) indicate that 
it has changed significantly in the past century.

  Motivated by  this background, we have obtained long--slit spectroscopy of 
IRC +10420 with the Space Telescope Imaging Spectrograph (HST/STIS), supplemented
by groundbased near-infrared spectra.  The high spatial resolution 
of STIS (0$\farcs$05 to 0$\farcs$1) allows us to separate the spectrum of 
the central star from the spectra of the surrounding ejecta for studies
of  morphology, velocity structure, and composition.  We find that
the spectrum at each location in the ejecta is  essentially that of a 
reflection or scattering nebula as expected, but some of the spectral 
details differ from those in 
the star.  The reflected spectrum gives us some novel information about the
stellar wind and the morphology of the ejecta.

  In the next section we describe our HST/STIS observations and 
groundbased near-infrared spectroscopy, and in Section 3  we describe the 
optical spectrum of the central source,  the lower resolution near-infrared 
data, and the subtle differences between the spectrum of the star and 
the inner ejecta.  Our analysis follows in the next sections. In section 4 we 
measure  velocities along the slit to assess the kinematics of the ejecta; we
find a surprising lack of variation of the reflected H$\alpha$ profile.  In 
section 5 we summarize the  implications for the central star and its 
wind, and the morphology of the ejecta.  We demonstrate that the wind 
of IRC+10420 is opaque or nearly so, and show how this  may account 
for the the mixed observations of both outflow and infall in its 
spectrum. In a final section we comment on the evolutionary state of 
IRC+10420 and compare it to the other well-studied yellow 
hypergiants in the context of de Jager's ``yellow void''.

\section{The Observations and Data Reduction}

  We obtained long-slit moderate resolution HST/STIS spectra of IRC +10420
and its ejecta  with a spatial sampling of 0$\farcs$05 
per pixel.  The effective resolution was
somewhat worse than 0$\farcs$05.  The 52{\arcsec} $\times$ 
0$\farcs$1 spectrograph slit was oriented along position angle 
58.3{\arcdeg} and located at two positions shown in Figure 1.  
Slit ``position 1'' was centered on the star and passed through
the bright inner ejecta, a small knot to the northeast, and 
across one of the arcs.   The slit centerline of ``position 2'' was
0$\farcs$42 from the star (measured perpendicular to the slit),
roughly southeast of position 1, and passed through a bright knot
east of the star and through the bright ejecta in the fan-like structure 
to the south and southwest of the star (see Paper II).  We used moderate
dispersion STIS gratings G430M and G750M at several tilts selected 
to cover the wavelength ranges 4560--5100 {\AA} and 6500--8600{\AA}.
The resulting spectral resolutions range from 8000 to 9000  and 5800 to 7600 
in the blue  and red data respectively.  These observations
were all obtained on 1999 September 10 and the journal of observations
is summarized in Table 1.

  We used the STIS Instrument Development Team's IDL-based data 
reduction package \citep{Lind99}, which includes the same basic 
steps as in the CALSTIS reduction package in STSDAS but has  
significantly  better distortion corrections
and rebinning techniques.  The spectra were flatfielded and flux and
wavelength calibrated. For the exposures at wavelengths longer 
than 7000 {\AA}, additional flat fields were obtained during 
Earth-occultation to help correct for fringing.  Cosmic rays were 
removed from data samples with multiple exposures, and corrections
for terrestrial motion were applied to all of the spectra.  

   The STIS IDT reduction package is superior to the
standard one for cases where high spatial resolution is desired, 
because it  minimizes the wavy flux-calibration errors that occur
in narrow extractions such as we use for the central star.
Since each STIS CCD row extends approximately along the dispersion
direction,  an extracted spectrum of a localized spatial region
conceptually amounts to the sum of a few CCD rows.  However, because 
the dispersion direction is not precisely
aligned with the rows, a ``3-row extraction,'' for instance, is
not merely one set of three rows across the detector.  The local
central row of the extraction must vary adaptively with column
number or wavelength, using interpolation techniques to combine
data from about four CCD  rows at each wavelength.   At some
wavelengths the extraction center is close to the middle
of a detector row, at others it is closer to the division
between two rows.  For point sources,  too simple a technique gives 
different results
for these two cases, leading to  variations in the calculated
flux level.  The resulting sinusoidal pattern in apparent flux
level can be  disagreeably large (several percent) in data processed
with the standard CALTSIS procedures, but is substantially less
when the IDT package is used instead.         

  We also corrected another, less familiar calibration error
that has usually been neglected in the past.  For the spectrum of
the central star we extracted an unusually narrow sample, only
3 rows (0$\farcs$15 wide), from each two-dimensional STIS spectral
image.  When such narrow extractions of STIS data are reduced
using the standard procedures, flux values for adjoining grating
tilts disagree by 5--10{\%} in the narrow wavelength overlap
sampled by both tilts.  This effect, which we discovered and assessed
earlier in data on $\eta$ Carinae,  indicates a consistent error
in the slope of the standard flux calibration for each grating tilt.  It is 
caused by a
variation in focus across the STIS detector, and becomes less
conspicuous for wider extraction samples, i.e., in cases where
poorer spatial resolution is acceptable.   We therefore applied
the appropriate  corrections to the spectrum slope for each grating tilt,
made minor corrections for small changes in pointing, and then
``spliced'' or merged the data from the two blue and four red
grating tilts.      

  Near-infrared spectra were observed with the Cryogenic Spectrometer
(CRSP) on the KPNO 2.1-meter telescope and with the PHOENIX high 
resolution infrared spectrometer on the 4-meter,  during June 2000 and
are summarized in Table 2.  The CRSP low-resolution spectra from 0.9 to 
2.5 $\micron$ (IJHK) were obtained using a 0$\farcs$8-wide slit  
aligned with the STIS slit at position angle 58$\arcdeg$. 
The I and H bands required 2 separate grating
tilts to cover the full wavelength range. The I and J band spectra were also
observed at an offset position 5$\arcsec$ south of the star at the
same position angle, but no significant differences were apparent there.
The seeing was typically $\sim$1$\arcsec$ on the
photometric nights and the spectra were extracted from a 4$\arcsec$--wide 
section of the slit centered on the central object.  The spectra
were sky subtracted by chopping along the slit, with offsets of 20 to
30$\arcsec$. Telluric absorption lines  were eliminated by dividing by 
the spectrum of HR 7396, after removing the hydrogen lines from the standard 
star's spectrum. The spectra were then flux calibrated using $\nu$ Her.     
Small corrections of the flux and wavelength were needed to merge adjacent 
spectral oders, similar to the adjustments made to STIS data as 
described earlier.  Since the CRSP data typically have a 2-pixel 
spectral resolution of several hundred km s$^{-1}$, no useful kinematic 
information is available from the near-infrared emission lines.

High-resolution PHOENIX observations of a few  bright
hydrogen emission lines were also made. The observing conditions were 
non-photometric, 
but IRC+10420 is bright enough that useful spectra were obtained with 
$\sim$1$\farcs$5 seeing. The 0$\farcs$6-wide slit was oriented at position angle
58$\arcdeg$, and sky subtraction was accomplished by chopping along
the slit, with offsets of 20 to 30$\arcsec$.  To remove telluric
absorption features, the spectrum of IRC+10420 was divided by the
spectrum of Vega.  Broad hydrogen absorption lines in Vega's spectrum
were easy to fit and remove at high spectral resolution, typically a
few to several km s$^{-1}$.

The details of the STIS spectroscopy of the central star and the inner
ejecta and the recent near-infrared spectra are described in the next 
section. While these results and the subtle differences between the spectra 
of the star and the ejecta are of interest and important for understanding 
IRC +10420's complex wind, its variability, and processes in the ejecta, 
the most novel results  and  implications for IRC +10420 are discussed 
in sections 4, 5 and 6.

\section{The Spectrum of the Central Star and the Inner Ejecta}

  The earliest spectrograms of IRC+10420 in the blue wavelength region 
showed the absorption line spectrum of a mid- to late F-type star of very 
high luminosity, with no obvious emission lines \citep{RMH73}.   
\citet{Irv86} was the first to report strong H$\alpha$ emission plus 
the [Ca~II] $\lambda$$\lambda$7293,7323 lines in emission;\footnote{
     HST/STIS spectra directly give vacuum wavelengths, but  
     we have chosen to quote the more familiar laboratory or 
     air wavelengths here.   Both forms are given in the tables 
     where appropriate.  All Doppler velocities quoted here
     are heliocentric.  For IRC +10420, 
     $V_{LSR} \, \approx \,  V_{hel} + 17.5$ km s$^{-1}$.}  
previous spectra of the H$\alpha$ region had shown no emission \citep{Fix81}. 
In Paper I we reported  strong emission in the Ca~II near--infrared 
triplet lines $\lambda$$\lambda$8498,8542,8662 and identified Fe~II and 
[Fe~II] emission lines in the red spectrum.  \citet{Oud98}  
published a high resolution echelle spectrum covering the 
wavelength range  3800--10400 {\AA}  and  identified many permitted 
and forbidden emission features, including some lines of relatively 
scarce elements such as  scandium, vanadium, yttrium, and potassium. 
None of the ground-based observations just 
mentioned had good enough spatial resolution to reliably separate 
the central star from its ejecta.

To examine the spectrum of the star and its wind with practically 
no contribution from the ejecta,  we extracted a spectrum just three 
rows wide in the STIS data, corresponding to $\sim$ 0$\farcs$15 
along the slit. The spatial sample was thus about 
0$\farcs$1 $\times$ 0$\farcs$15, corresponding to a 
projected size of 500 $\times$ 750 AU at the nominal distance 
of 5 kpc.  From these data we deduced the star's total apparent 
flux by applying suitable throughput corrections.   These included 
some factors omitted in most other STIS work, noted in Section 
2 above.

  Figure 2 shows the resulting flux--calibrated spectrum  in our blue and red 
wavelength ranges.  Uncertainties in absolute flux values are 
difficult to assess because they are mainly systematic, arising 
in the data reduction and calibration process;  
$\pm$ 5{\%} is a realistic  estimate.  Figure 2 
also shows a satisfactory fit to the continuum, a simple parabola 
in this log-log plot.  If we measure flux and wavelength in 
erg cm$^{-2}$ s$^{-1}$ {\AA}$^{-1}$ and {\AA}, then this smooth
continuum is specified by 
${\log} \; F_{\lambda}  =  -13.701, -12.925, -12.432$   at
${\log} \;  {\lambda}  = \ 3.7, 3.8, 3.9$  respectively ($\lambda$ 
= 5012, 6310, 7943 {\AA}).   Since it matches the observed values 
within the likely errors, there is no need to seek a more 
sophisticated fit at this time.    In Figure 3,  which shows the 
relative spectrum ``normalized'' to have continuum levels close 
to unity, we divided the observed spectrum by this simple 
log-log-parabolic continuum (rather than by a local 
continuum fit which would be less well defined). 

\subsection{The Energy Distribution of the Central Object}

   Available groundbased photometry and spectroscopy of IRC +10420
cover regions 1{\arcsec}--10{\arcsec} across,  and therefore 
include various fractions of the circumstellar nebula.  In Paper II 
using the HST/WFPC2(PC) images we found that the central object accounts 
for less than 50{\%} of the total apparent blue and visual-wavelength 
light.  As one might expect for a reflection nebula, the  ejecta appear 
bluer than the star.  We have used our STIS data (Fig.\ 2) to estimate
apparent magnitudes of the star over a wider wavelength range.
As mentioned above, we sampled a 0.015-square-arcsec area and then 
applied suitable throughput corrections.  For wavelengths not 
included in the observations we adopt the smooth continuum fit 
shown in Fig.\ 2. The results are listed in Table 3;  likely errors
are mostly systematic, informally about $\pm$0.05 for 
each magnitude and  for each difference (color).
Groundbased  observations show that IRC +10420 slowly brightened 
about a magnitude in the blue  from 1926 to
1970 \citep{Got78}, but its visual flux (Paper I) has  remained essentially constant 
since then. Our more recent space-based measurements are consistent 
with little or no change since the last ground-based photometry in 1991. 

   IRC+10420 is a  highly reddened object due both to its large distance 
and to an  uncertain amount of additional circumstellar reddening. 
Assuming intrinsic colors appropriate for a late A to early F--type 
supergiant, the observed colors for the central star (Table 3) correspond 
to a visual extinction  A$_{V}$ $\approx$ 6.5--7 mag. In Paper I 
we showed that its high polarization in the visual is interstellar, 
corresponding to a high interstellar extinction of  6--7 mag thus 
accounting for the observed color excess. However its large infrared excess
indicates a large amount of circumstellar dust which should form at 
roughly 100 AU (0$\farcs$02) from the star. In Paper II we found that 
the visual optical depth $\tau$ due to absorption and scattering is 
about unity at  $r$ $\approx$ 1$\arcsec$ (5000 AU). Thus we would 
expect $\tau \, \approx \, 50$ at 100 AU and the star should be 
invisible. To reconcile these inconsistencies, we suggested that 
either the dust is ``patchy'' or inhomogeneous, or else that we view 
the star's photosphere near the polar axis with only a modest amount 
of circumstellar extinction along our line of sight.  We will return 
to questions concerning the geometry of IRC+10420's circumstellar 
ejecta in sections 4 and 5. 

\vspace{2mm}

\subsection{The Emission and Absorption Line Spectrum of the Central Object} 

IRC +10420 has a  complex spectrum of absorption lines and permitted and
forbidden emission lines  dominated by strong emission from  H$\alpha$, 
H$\beta$, the Ca II infrared triplet, and [Ca II], as shown in Figure 3. 
The H$\alpha$, H$\beta$, and permitted Ca II lines all have prominent 
double-peaked emission profiles with a stronger  blue component.  Figures 
4 and 5 show these profiles for the central star from the  STIS data.   
The measured velocity separation  of the  emission peaks for all four 
lines is 100 to 120 km s$^{-1}$ with the central absorption at 
+ 71.2 $\pm$ 1.3 km s$^{-1}$ (heliocentric).\footnote
    {The systemic velocity is usually assumed to be 
    +58--60 km s$^{-1}$ (heliocentric), based on the OH and CO 
    observations \citep{Bow84, Knapp85, Oud96}} 
The velocities of the emission peaks with respect to the central 
absorption imply outflow or wind speeds of 50 to 60 km s$^{-1}$, 
somewhat higher than the expansion velocity of $\approx$ 40 km s$^{-1}$ 
inferred from the OH maser  and CO observations \citep{Bow84, Knapp85}.
In Paper I we attributed the double  emission  peaks to rotation 
of a hypothetical disk or ring, while  \citet{Oud94} suggested
they might be due to a
bipolar outflow. However, in Section 4 below we report a 
serious difficulty for either of these interpretations;   
these profiles most likely result, instead, from radiative transfer effects 
in the star's optically thick wind. The  measured velocities for 
these lines are summarized in Table 4.

We first  noted the extraordinarily broad wings of the 
H$\alpha$ profile in Paper I. In the STIS data for the central star they are easily visible out to more
than $\pm$ 1000 km s$^{-1}$ for both H$\alpha$ and H$\beta$ (Fig.\ 4).  
In the relatively low-speed wind of IRC+10420,  the most obvious
explanation for these wings is scattering by free electrons.   
Their relative strength is reminiscent of $\eta$ Car, suggesting 
that ${\tau}_e \ {\sim} \ 1$ in the region where the escaping Balmer
photons originate (cf.\ Davidson et al.\ 1995, Hillier et al.\ 2001). 
In principle this reasoning can produce a new estimate of the 
mass loss rate, but calculations beyond the scope of this paper   
would be required.  For that purpose one must know the intrinsic H$\alpha$ 
luminosity corrected for extinction, and other details.

Similar broad wings may be present on the Ca~II lines.  For example, 
Ca~II $\lambda$ 8542 has a prominent blue wing (Fig.\ 5) which resembles 
H$\alpha$, while the red side is masked by Paschen absorption.  The 
weaker Ca~II $\lambda$8498 line may also have a faint blue-side wing, 
and the third member of the triplet, $\lambda$8662, was not included 
in our STIS observations.  We also note that $\lambda$8498 is narrower 
than either of the brighter lines, $\lambda$8542 and H$\alpha$, and 
this may  have some significance for models.

The Ca II emission lines are probably produced in the star's
wind or envelope by radiative de-excitation from the strong Ca~II H 
and K absorption upper levels. The [Ca II] emission lines however are 
not often seen in stellar spectra; their upper states would  
normally be collisionally de-excited back to the ground state. In Paper I 
we showed that about one-fourth of the excitations  result in the 
observed [Ca~II] lines and the electron density is roughly three 
times higher than the critical density. We have repeated that  calculation 
with our high spatial resolution STIS data and confirm our previous 
estimate.  The [Ca II] lines are thus produced in a region of much 
lower density than normally found in the atmospheres of even the 
most luminous normal  A and F--type supergiants.  Presumably these 
lines originate outside the dense the wind, at $r\; >\; 2R_{star}$ where
$n_e \; <\; 10^8$ (see section 5 below).

In addition to the specific lines discussed above,  numerous permitted
and forbidden emission lines of ionized metals, including  some relatively scarce 
elements,  dominate the red 
spectra:  Fe~II, Ti~II, Sc~II, Y~II, Cr~II and 
[Fe~II] and [V~II].  Most of the identified absorption lines of the 
ionized metals  such as Fe~II, Ti~II, Cr~II and Sc~II plus others are observed 
at blue wavelengths, but strong absorption lines of O~I, N~I and 
the Paschen lines are easily visible in the far red specta (Fig.\ 3). 
The general appearance and relative strengths of the emission and
absorption lines in our STIS spectrum of the central star are consistent
with the echelle spectrum published by \citet{Oud98}.
Our measured velocities for the emission and absorption lines
are summarized in Tables 4 and  5, and agree well with the velocities for  
the same lines reported by \citet{Klock97} and \citet{Oud98}. Thus the 
STIS data provide confirmation that groundbased results were not 
misleading due to contamination by the ejecta.

{\it The Spectral Type:} The first low resolution spectrogram of 
IRC +10420 showed the  features of a very luminous mid- to late 
F--type star, with a G band and strong Ca~II H and K lines 
\citep{RMH73}. However, the absorption lines in 
recent higher resolution digital spectra indicate a much
earlier spectral type and therefore a warmer apparent temperature. 
\citet{Oud98} has compared the strengths of the absorption lines in the  
wavelength range 4690--4935 {\AA} in his 1994 echelle spectrum to 
several supergiants with spectral types ranging  from A2 to F8, and 
concludes that the lines are most consistent with an F0 to A2 type 
and may even be earlier than  A2. Both the appearance and equivalent 
widths of the absorption lines in our STIS data agree with Oudmaijer's  
measurements.  Thus there does not appear to have been any major 
change in the 5 years between the two spectra.  We caution, however, 
that a spectral type inferred from a high-resolution echelle  spectrum 
is not the same thing as an MK spectral type based on a low-resolution 
photographic image tube spectrum. Since many of the features used in
the low resolution  MK classification are resolved in the echelle spectrum,
these  spectral types are not based on the same features.  In this 
connection, however, \citet{Klock97} estimated a temperature of 
8500 $\arcdeg$K, appropriate for a middle A-type supergiant, based on  
a simple atmosphere analysis of spectra  from 1992--96. 
Furthermore, the equivalent widths of the absorption lines
plus the presence of a weak G band, and the Ca I $\lambda$4226 line in a 
moderate resolution echelette spectrum from 1988  (Paper I) 
indicate a late A to  early F-type star of high luminosity. 
So it appears that the apparent spectral type of IRC +10420 changed 
from a mid/late F-type to as early as a middle A type  between 1970 
and 1990. This change in spectral type and by  implication a corresponding 
temperature increase,  may not be due to evolution, but may instead 
reflect variations in its wind. This possibility is discussed more 
fully in section 5.                

{\it Line Identifications:} \citet{Oud98} gives an exhaustive line 
identification list in his on-line table. We confirm his identification 
of Fe~I (multiplet 60) in emission, but not the three [Fe~I] (multiplet 14)
lines  at $\lambda\lambda$7172,7388,7452. 
Instead we suggest that these are [Fe~II] (multiplet 14) and that the 
entry in his table is a typographical error. He has also identified 
[O~I] emission at $\lambda\lambda$6300,6363, in a wavelength region not 
covered by our STIS spectra. Additional absorption lines in his table 
in those other wavelength regions include Y~II, V~II, Sr~II, Si~II, Mg~II 
and some neutral metals, Si~I, Mg~I and possibly even He~I $\lambda$5876;
however, no other He I  lines are observed including $\lambda$10830,
see section 3.5.  

Although we see most of the lines identified by Oudmaijer in the wavelength
ranges that we observed,  some lines in our STIS spectra are either not apparent 
in his groundbased 
spectra or  were not included in his table of identifications. 
In Table 6 we list some possible additional line identifications with their 
measured equivalent widths and velocities,  and in Table 7  we list 
the unidentified lines in our STIS spectra. We used the revised multiplet
line identification table \citep{Col93} available at the Astrophysics
Data Center (Catalog 6071A)  and the emission line identification list for 
$\eta$ Carinae
by \citet{Zeth01}. We calculated the vacuum  and  air  wavelengths for the
lines in Tables 6 and 7  using our mean velocities  of the absorption 
lines of the ionized metals  (86.2 $\pm$ 8.5 km s$^{-1}$) and of the 
permitted and forbidden emission lines (66.6 $\pm$ 8 km s$^{-1}$) in Table 5. 
With $\sim$ 8 km s$^{-1}$ rms scatter about
the mean velocity,  the calculated  wavelengths may be uncertain by as much as 
0.2 {\AA}. Some of these  unidentified lines  
are potentially interesting, and several are relatively strong,
such as the two lines near 7000 {\AA}.

{\it Inverse P Cygni Profiles:} \   
\citet{Oud95} reported that many of the ionized metal lines have emission
and absorption on their {\it blue\/} and {\it red\/} sides respectively,
but not all of the ionized metallic lines have profiles of this type.   
We confirm the presence of inverse P Cyg profiles in wavelength regions 
sampled both by our data and by Oudmaijer's,  although in some cases either 
the emission or absorption is doubtful in our spectra.  Our measured 
mean velocities are +40 $\pm$ 7 km s$^{-1}$ and +107 $\pm$ 7 km s$^{-1}$ 
for emission and absorption components, respectively. The velocities 
of absorption components differ by about 20 km s$^{-1}$ from Oudmaijer's 
(1998) published velocities for the same lines, in contrast with the
overall good agreement for ``normal'' absorption and emission lines. 
\citet{Oud02} also finds variability in the velocities from the inverse 
P Cygni profiles. This variability might be due to the changes 
in the relative strengths of the emission and absorption components, 
although in our spectra the emission components are rather weak in 
most of the profiles and the absorption profiles are usually symmetric,
suggesting that they have not been distorted by the emission. Therefore we
suspect that the velocity changes indicate variability in the outflow and
infall of the material. In our data the inverse P Cygni 
emission  component is blueshifted 27 km s$^{-1}$ with respect to the 
mean velocity from the normal ionized metallic emission lines, consistent with 
the interpretation that these lines are due to infalling gas. In 
contrast the absorption 
component has a velocity 20 km s$^{-1}$ larger than or redshifted with 
respect to the mean velocity of the ionized metal absorption
lines in the rest of the spectrum (see Table 5). 

In section 5 we discuss 
the kinematics of IRC+10420's wind and ejecta and propose a model that 
accounts for the apparent contradiction of both infall and outflow.  

\subsection{Spectra of the Bright Inner Ejecta}

IRC+10420's circumstellar nebula is a reflection nebula. However, when
we examined the spectra of the brightest ejecta, within 1$\arcsec$ of the
star, we did notice some spectral differences which  may  
be useful for understanding  the processes in IRC+10420's  wind
and its immediately surrounding ejecta.

The central source is so  bright  at our longest wavelengths that photon 
scattering in the STIS CCD degrades the  spectra of the ejecta in  slit 
position 1.  This scattered light problem is most obvious  longwards of 
8000 {\AA}, but would compromise any equivalent width measurements for 
the ejecta beyond 7000 {\AA}  at this slit position.  For this reason, 
we restrict our discussion of the ejecta spectrum to two locations on 
slit position 2: ($\underline{a}$) the bright ejecta in the 
``fan'', 0$\farcs$45 south of the star, and ($\underline{b}$) the 
bright knot to the east of the star (see Figures 1 and 12). 

The spectra on the star and at the two offset positions are essentially 
the same except for a few absorption lines and an emission line at position 
$\underline{a}$ for which the differences are quite pronounced. The 
individual lines  with their wavelengths, equivalent widths and atomic 
transitions are listed in Table 8. Four of the absorption lines, Fe~II 
(multiplet 42) at $\lambda$$\lambda$4924 and 5018{\AA}, Cr~II (multiplet 
30) at $\lambda$ 4824{\AA} and a strong unidentified line measured at 
$\lambda$6963{\AA} (Table 7) are much stronger in the star with 
equivalent widths more than three times greater than in the reflection 
spectrum. The third line in the Fe~II (42) series is just outside our 
spectral coverage. The other lines in Cr~II multiplet 30 observed in 
our spectra are also stronger in the star's spectrum, but the effect 
is most pronounced for the strongest line in the series.  Close 
inspection of the line profiles shows a raised continuum on either side 
of the these absorption lines, suggesting that emission may be 
responsible for the weaker absorption lines in the ejecta.  One emission 
feature, the unidentified line measured at $\lambda$8105 {\AA} (Table 7) 
is stronger  at position $a$ than in the star. The spectrum at position 
$\underline{b}$ does not show these differences and seems essentially
identical to that of the star although the S/N is poor.

  The behavior of the emission lines of K~I at $\lambda$$\lambda$7665 and 
7699 {\AA} identified by \citet{Oud95} in his echelle spectrum with 
inverse P Cygni profiles is more subtle in our spectra.   K~I emission 
is not apparent in our STIS spectrum on the star itself (Figure 6), 
although the strong interstellar K~I absorption lines and weak stellar 
absorption components are present.  However, in the spectrum of ejecta
at location $\underline{a}$,  $0\farcs$45 from the star, K~I emission 
is clearly present and easily seen above the continuum (Fig.\ 6). 
We are confident of the identification because our measured velocities 
for the three K I components ( IS abs., em., stellar abs.) agree with  
Oudmaijer's groundbased measurements and the K~I emission line velocity 
also agrees with the velocities for the other emission lines in 
IRC +10420 (Table 5).  

Strong  K~I emission has previously been observed in the extreme red 
supergiant VY CMa \citep{Wall58}, and weaker emission has been detected 
in several stars including  the circumstellar gas shells of the M--type
supergiants $\alpha$ Ori \citep{Ber75,Ber76}, and  $\mu$ Cep \citep{Mau97}, 
and in the spectrum of the hypergiant star $\rho$ Cas \citep{Lob97}.
K~I has a very low ionization potential of only 4 eV. and is expected to 
form in low temperature, low density regions.  Wallerstein's (1958) 
comparison  of the intensities of K~I and Na I emission lines in 
VY CMa yielded a temperature of only about 700 $\arcdeg$K for the 
region where the lines are formed, assuming they come from the same gas. 
We therefore suspect that this rare line is formed  outside the dense 
wind of IRC +10420, e.g., 0$\farcs$45 or 2000 AU from the star as shown
in Fig.\ 6.   We therefore looked for K~I emission at different locations 
along slit position 1, and found that the K~I emission components 
become clearly visible above the level of the continuum at distances 
of 0$\farcs$2 and 0$\farcs$5 from the central star to the NE and SW 
respectively. Thus it appears that  the K~I emission is formed in 
circumstellar gas at distances of a 1000 AU and greater from the star. 
This is not unlike $\alpha$ Ori whose K~I emission extends from 
5$\arcsec$ to as far as 50$\arcsec$ or 700 to 7000 AU \citep{Hon80}.

\subsection{The Diffuse Interstellar Bands}

Given the large distance, it is not surprising that there are many diffuse 
interstellar bands (``DIB's'') in the spectrum of IRC+10420. \citet{Oud98} 
measured several bands and derived an approximate color excess ($E_{B-V}$)
attributed to interstellar dust. We thought it worthwhile to measure
the strengths of these bands in our spectra on and off the central star to
look for any differences that might be attributed to spatial variation across
the line of sight or to additional circumstellar reddening of the star. 
Using the $W_{\lambda}$/$E_{B-V}$ ratios in \citet{Jenn94} we find mean 
$E_{B-V}$ values of 1.84 $\pm$ 0.81 and 1.87 $\pm$ 1.3, respectively, 
for 13 DIB's in the stellar spectrum and for 11  at the position 
0$\farcs$45 south of the star.  We attribute the large error at the 
offset position to the lower S/N in the spectrum.  These results are 
somewhat higher than Oudmaijer reported and imply an A$_{v}$ $\sim$ 6.1 mag, 
slightly lower than that inferred from the observed colors of IRC+10420 
(see section 3.1). This does not necessarily mean that the difference
is due to circumstellar reddening. The strength of the DIB's indicate 
extinction due to the diffuse interstellar medium and do not include
contributions from molecular clouds along the line of sight, which can 
be significant for an object at the distance of IRC+10420. These results 
are therefore consistent with our earlier conclusion that most of the 
observed reddening of IRC+10420 is interstellar.

\subsection{The Near--Infrared Spectrum}

The  low-resolution near-infrared spectrum of  IRC+10420 in Figure 7  
is  dominated by emission lines of the Paschen and Brackett series of
hydrogen and lines of low-ionization metals like Na~I, Fe~II, and Mg~I. 
We see some significant differences compared to earlier spectra obtained 
at the same wavelengths.  The lines of Na~I and Mg~I noted by 
\citet{Thom77} are all present in our spectra, plus the prominent 
Mg~I line at 1.488 $\micron$ which was absent in their 1976 spectrum.  
The absence of this line was used to justify their suggestion that the 
Mg~I and Na~I emission was due to
pumping by blue and UV photons from an optically thick region, 
a possible chromosphere.  Numerous Fe~II and [Fe~II] lines are now observed 
while none were reported by \citet{Thom77}, and  the Fe I line that they  
detected at 1.601 $\micron$ is absent in our more recent spectrum.
\citet{Fix87} identified lines of Fe II in their high dispersion spectra 
from June 1984 with  intensities comparable to what we observe.   
It appears that the most significant changes in the strength of
low-ionization metal lines in IRC+10420's near-infrared spectrum occurred
between 1976 and 1984, consistent with the changes in apparent
spectral type observed at optical wavelengths. Our line identifications
are in Table 9.

CO bandhead emission at $\sim$2.3 $\micron$ may be present in Figure
7, but the spectra are somewhat noisy and these wavelengths overlap
with the higher-order Pfund series of hydrogen.  Higher quality
spectra are needed to further investigate these lines.  No hint of
molecular hydrogen emission at 2.122 $\micron$ is seen in our spectra.
Also, no He I lines are seen in Figure 7, which is relevant to changes
in IRC+10420's spectral type and characteristic temperature described earlier.

Atomic hydrogen lines in IRC+10420's near-infrared spectrum have also 
shown significant variability.  Early near-infrared spectra by 
\citet{Thom77} and \citet{Fix87} showed the  hydrogen lines in
absorption.  \citet{Oud94}  later discovered  Br$\alpha$, Br$\gamma$, 
and Pf$\gamma$ in emission with the  higher-order Brackett (i.e. Br 12 
and Pa 11) in absorption in the mid-1990's. Our spectra confirm these
results  and also show the Paschen lines  in emission up to Pa 9.  
\citet{Oud94} noted  that  the near-infrared hydrogen  emission 
line profiles were asymmetric,  the Br$\gamma$ and Pf$\gamma$ 
showed  inverse P Cygni profiles, and the 
redshifted emission peak seen in H$\alpha$ was missing in these 
hydrogen lines.  Our high-resolution spectra of a few H line profiles
in Figure 8 do not show inverse P Cygni profiles, but are asymmetric; 
the emission peaks of all these lines are blueshifted.  Fits  to the 
line profiles did not yield satisfactory estimates of the centroid velocities 
because of the  significant asymmetry.  Instead, we measured the heliocentric 
velocity of each line by bisecting it at half the maximum intensity; the 
mean velocity is +36 $\pm$ 2.6  km s$^{-1}$, blueshifted by 
$\sim$ 22 km s$^{-1}$ with respect to the presumed systemic velocity,  
similar  to the results found by \citet{Oud94}. The velocity of the 
Br$\gamma$ is about 5 km s$^{-1}$  higher than the velocities for Br$\alpha$ 
and Pa $\beta$ indicating that it may be formed deeper in the wind. 

Finally, Figure 8 confirms the {\it absence} of He I $\lambda$10830 in the 
near-infrared spectrum.  He I $\lambda$10830 should be the strongest line 
of this species if it is present, so its  absence  raises some doubt about 
the identification of the feature near $\lambda$5876 as He~I 
\citep{Oud95,Klock97}.  The non-detection of He I $\lambda$10830 at this
time may be useful as a future benchmark if IRC+10420 continues to get 
warmer.  This line was detected in other objects observed the same night 
and with the same grating tilt configuration.

\section{Geometry of the Inner Ejecta and the H$\alpha$ Profile}

   Aside from a few differences noted above, the ejecta show 
the spectrum of the star reflected by dust.  Therefore, in effect 
{\it we can observe the star's spectrum from various directions.\/}  
This is potentially valuable because most models for IRC+10420 have 
invoked a circumstellar disk, or bipolar outflow, or other departures 
from spherical symmetry, which can be tested by inspecting line 
profiles as functions of position along the slits. 
   
   To be confident that the observed locations represent 
a wide range of directions from the star, we need information 
about the three-dimensional distribution of the scattering material.
Doppler shifts of the reflected spectrum can serve this purpose in
a way explained below.   The most suitable reference feature is the 
minimum in the bright double-peaked H$\alpha$ profile.  Figure 9 
shows velocities of this feature measured at various locations;  
we used extractions 3 pixels wide for the inner samples,  
5 pixels wide at radii between 1$\arcsec$ and 1$\farcs$5, and 
10 pixels wide beyond that.  Due to expansion (see below), apparent 
Doppler shifts tend to increase in each direction away from the star;  
two straight lines in the figure conveniently mark these trends 
for slit position 1.  

A brief analysis shows the relation between  apparent velocity and 
three--dimensional position. 
The apparent Doppler velocity seen in a dusty reflective 
condensation is  \begin{displaymath}
      V_{obs}  =  V_{0} + V_{r} + V_{z} ,
\end{displaymath}
where  $V_{0} \; \approx \; +74$ km s$^{-1}$ is the value for the
same spectral feature when the star's spectrum is observed directly,  
$V_r$ is the outward velocity of the condensation away from the star, 
and $V_z$ is the corresponding relative velocity component along our 
line of sight.  For demonstration purposes, adopt some constant ejection 
speed $V_{r}$ and consider only slit position 1 which includes the star.   
Let $x, z$ be spatial coordinates centered on the star; 
the $z$-axis extends along our line of sight ($z$ increases with 
distance) while the $x$-axis is parallel to the spectrograph slit 
($x$ increases toward the northeast).  For a given observed location, 
we know $x$ from its position along the slit and we calculate 
$V_{z}  =  V_{obs} - V_{0} - V_{r}$. 
Then $V_x = {\pm} \sqrt{{V_r}^2 - {V_z}^2}$,  which allows 
us to deduce $z = (V_{z}/V_{x}) x$.   
(See \citet{KD01} for a more detailed application of this method
to another object.)

   Figure 10 shows a set of illustrative results 
in the $xz$ plane, using three sample ejection 
speeds $V_r$ = 25, 50, and 100 km s$^{-1}$.  Here 
our viewpoint is to the left of the figure, and each set of two curves 
(northeast and southwest) shows the spatial loci corresponding to 
the linear velocity trends drawn in Figure 9.   We expect to 
see mainly the near side of the nebula, since the optical 
thickness for visual-wavelength internal extinction by dust 
exceeds unity out to a radius of about 2{\arcsec} (Paper II);
therefore the curves in Figure 10 seem qualitatively reasonable except 
near the center where they are not expected to be valid 
(cf.\ Fig.\ 9).  For assumed values of $V_r$ less than about 
30 km s$^{-1}$, observed material on the northeast side extends 
implausibly far along the line of sight;  
while values greater than 80 km s$^{-1}$ entail a similar difficulty 
on the southwest side.  In other words, {\it the structure has a 
likely shape if $V_r$ is between, roughly, 35 and 70 km s$^{-1}$.}   
This statement is independent of arguments based on line widths 
in the stellar spectrum (Section 3 above) and is consistent with the 
outflow velocities inferred from the CO and OH observations and from 
the separations of the hydrogen and Ca II double emission peaks.  

   Figure 11 shows the spatial distribution of the data points 
in Figure 9, assuming that $V_r$ = 50 km s$^{-1}$.  The uncertainties
are too large for this to be a formal model, but Figs.\ 10 and 11 
confirm two expected but crucial generalities:  (1) The Doppler 
velocities  are consistent with reflection in an expanding
nebula of the type envisioned in the past (e.g., in Paper II);  
and (2) the observed locations ``see'' the star from a wide range 
of directions, spanning at least a 90{\arcdeg} angle.  Slit position 
2 further widens the range of viewpoints. Moreover, the southwest 
side appears to be  a little closer to us than 
the northeast side, and  if the configuration has bipolar symmetry to
some extent, then its axis is most likely tilted roughly 65{\arcdeg} 
relative to our line of sight, consistent with the model 
we proposed in Paper II.

    At least one smaller-scale structure is obvious in Figs.\ 9 
and 11:  the apparent Doppler velocity trend is locally reversed 
between +1$\farcs$2 and +2{\arcsec} NE,  where slit position 1 
crosses one of the arcs reported in Paper II. To investigate this 
more thoroughly we measured 3-pixel-wide samples (the hollow squares 
in the figures) and the trend remains consistent at that narrower 
sampling interval.  The resulting feature in Fig.\ 11 probably 
indicates that the arc structure is more or less flat rather than 
jet-like or shell-like, and indirectly represents the spatial 
orientation of a cross-section.

\subsection{The Surprisingly Uniform H$\alpha$ Profile}

  We have measured the H$\alpha$ profile at four locations along
slit position 1 and two along position 2, shown in Figure 12; this 
line is bright enough to provide adequate signal/noise even at 
places where other spectral features are uncertain.  Figure 13 
shows the results along with the star's apparent H$\alpha$ profile.
The shape of the double-peaked profile is practically the same at 
every location, if we allow for some blurring by local velocity 
dispersions in the reflecting material.  Most significantly, within 
the uncertainties,  {\it at every location the difference between 
the two peaks agrees with the 122 $\pm$ 10 km s$^{-1}$  measured 
for  the H$\alpha$ profile in the stellar spectrum.\/}
This result conflicts with most proposed explanations for the 
double-peaked profile.  For instance, if the peaks represent
rotation in a circumstellar disk or torus as we suggested in
Paper I, then their projected velocity difference should include 
a factor of $\sin \, {\theta}$ where $\theta$ is the angle between 
the disk axis and the local viewing direction.   Since the viewing 
direction varies considerably among the sampled locations, that 
model predicts an obvious variation contradicted by Figure 13.  
The same difficulty arises, perhaps even more so, for the polar 
outflow model suggested by \citet{Oud94}.  In order to reconcile
the observations with either type of axial model, one must hypothesize 
a geometrical distribution of observed points that seems unlikely 
in light of the preceding subsection above.  {\it Therefore we conclude 
that the double-peaked line shape arises in a roughly spherical 
arrangement of circumstellar gas,\/} rather than an axially symmetric 
or seriously asymmetric configuration. In Section 5 we comment on 
likely reasons for the double peaked shape.

\section{The Physical State of IRC +10420's Wind or Envelope}

  In this section we review the physical context for spectroscopy 
of IRC +10420, and we propose several new hypothetical phenomena.  
Two unusual circumstances are crucial:  (1) the outer layers 
of this object exemplify a form of the modified Eddington limit 
\citep{HD84,HD94} and (2) as a consequence, its dense wind is 
probably opaque.  Recent discussions by \citet{deJ97b,deJ98,deJ01}
and \citet{Lob01} are broadly consistent 
with the views expressed below, but those authors were 
primarily concerned with less extreme objects such as
$\rho$ Cas and HR 8752, and hence did not emphasize the 
distinctions between a nearly-static photosphere and an 
opaque wind.  Since we approach the topic from a different 
viewpoint, our emphasis and terminology differ from theirs 
and so do some of the physical processes.

\subsection{An Opaque Wind}

   \citet{KD87} assessed the criterion for a stellar wind
to be opaque in the continuum and, if it is opaque, the resulting
photospheric temperature as a function of luminosity and mass-loss 
rate.  The temperature range 6000--8500 K observed for IRC +10420 
is rather special:   Due to the behavior of opacity, 
{\it opaque wind photospheres automatically 
fall within this temperature interval for a wide range of mass 
loss rates.\/}\footnote{
    This statement applies also to supernova photospheres and 
    LBV or S Doradus eruptions.  In a sense it plays the same 
    role for winds that the Hayashi limit does for convective 
    atmospheres.} 
Given the usual estimates for IRC +10420's mass loss rate 
and wind velocity quoted in Section 1 
above, its wind density parameter $Q$ defined in \cite{KD87} 
is of the order of $10^{-4.8}$.  According to Fig.\ 1 of that 
paper, this is large enough to make the wind opaque and to 
give it a photospheric temperature in the critical range.  

   Therefore {\it traditional static atmosphere models and spectral
classification cannot safely be applied to this object.\/}  Conventional 
''effective temperature'' is ill-defined in this case and should
be replaced by some other form of characteristic temperature 
\citep{KD87,HD94}.
In principle the underlying star may have evolved farther ``blueward'' 
than the A--F spectral type suggests, since the observable 
features originate in the wind.   {\it Spectroscopic variations 
observed in the opaque wind during the past 30 years represent changes 
in the wind parameters, and therefore do not necessarily indicate 
major evolution of the stellar radius and interior in that time.\/}   
(Proviso:  These assertions assume that the current mass loss rate 
is at least $1.5 \times 10^{-4}$ $M_{\odot}$ yr$^{-1}$, based on the 
published estimates which mostly represent long-term average values. 
IRC +10420's extreme time-averaged mass loss rate distinguishes it 
from the other two famous yellow hypergiants, $\rho$ Cas and HR 8752.)  

  Emergent radiation at any particular wavelength is created 
at locations where 
$ \sqrt{3 ({\tau}_{sc} + {\tau}_{abs}) {\tau}_{abs}} \; {\sim} \; 1, $
if ${\tau}_{abs}$ and ${\tau}_{sc}$ are optical depths 
for absorption and for scattering by free electrons.  
Idealized pseudo-LTE calculations with 
$v_{wind} \; {\approx} \; 30$ to $60$ km s$^{-1}$ provide a fair 
working model -- merely an initial guess -- for the likely state 
of IRC +10420.  The wind is translucent,  
${\tau}_{sc} \; {\approx} \; 1$, at radius 
$R \; {\approx} \; 1.6 {\times} 10^{13}$ cm $\approx$ 1 AU, where
$T \; {\approx} \; 9000$ K and $n_e \; {\sim} \; 10^{12}$ cm$^{-3}$. 
The Balmer continuum originates in this region because 
${\tau}_{abs} \, > \, 0.2 {\tau}_{sc}$ near the Balmer edge.  
At longer wavelengths ${\tau}_{abs} \, < \, 0.05 {\tau}_{sc}$, 
so the emergent visual-wavelength continuum must be created in deeper 
layers.  The ``stellar surface'' is best defined as the sonic point 
in the flow (cf.\ Davidson 1987, deJager 1998),  
presumably at some temperature above 9000 K.  Absorption 
lines, on the other hand, may be formed around 
$R \; {\approx} \; 2.2 {\times} 10^{13}$ cm $\approx$ 1.5 AU 
where ${\tau}_{sc} \; {\sim} \; 0.1$,  $T \; {\sim} \; 7000$ K, 
and $n_e \; {\sim} \; 10^{11.2}$ cm$^{-3}$.  The energy densities
of thermal gas, bulk motion, and radiation there are roughly 
0.6, 6, and 10 erg cm$^{-3}$ respectively;  and, given the likely
turbulent state in lower layers, we should not be surprised if the 
magnetic energy density is of the order of 0.1 erg cm$^{-3}$ or 
conceivably larger.  Outside this radius the wind becomes cool and 
transparent as the hydrogen recombines.   This model is too simplified 
to quote in more detail, but it evokes the basic nature of the case.  
A proper self-consistent NLTE model would be far beyond the scope 
of this paper, especially since three-dimensional gasdynamic 
calculations are required as noted later.

   The chief reason for the opaque wind is that this 
star's $L/M$ ratio is dangerously close to the Eddington limit,
in a non-trivial sense described below.  Other factors help 
destabilize the outer layers, but $L/M$ is of paramount 
importance.  A good way to approach the problem is to combine 
various remarks made by \citet{HD94}, \citet{deJ98}, and 
\citet{Lob01}, plus a few additional ideas. 

   Consider the probable evolution of a star with initial mass around 
40 $M_{\odot}$.  Despite substantial mass loss it can evolve to become 
a red supergiant.  (30--50\% more initial mass would entail LBV-like 
behavior that precludes the RSG stage.)  Suppose that the red supergiant 
then loses enough mass to raise $L/M$ above the classical Eddington 
limit.  This does not cause immediate trouble, because the atmospheric 
opacity at low temperatures is far less than the Thomson scattering 
value normally applied in  
$(L/M)_{Edd} \; = \; 4 {\pi} c G / {\kappa}$.
(Violating this limit in the interior merely ensures vigorous
convection.)  However,  when the star later evolves blueward away 
from the RSG part of the H-R diagram, ionization and opacity 
near the surface increase rapidly as the photospheric temperature 
rises above 6500 K;  then the Eddington limit becomes applicable,
other effects help to destabilize the atmosphere, and a terrific 
wind ensues.  The star has reached de Jager's (1998) ``yellow void'' 
in the H-R diagram.  

   The {\it modified\/} Eddington limit is relevant here, i.e.,  
temperature-dependent absorption must be included in the opacity 
and the absorption peak between 8000 and 12000 K decreases the 
limiting value of $L/M$.  This assertion is more subtle than it 
appears at first sight.  Humphreys and Davidson (1984, 1994) and 
\citet{deJ98} recognized that absorption scarcely affects the 
traditional static-atmosphere Eddington limit,  because atmospheric 
density decreases as the limit is approached, reducing the 
absorption opacity so that asymptotically one is left with 
only Thomson scattering.  Hence the ``modified Eddington limit'' 
is really a hypothesis  that instability occurs if $L/M$ exceeds,
say, 80 or 90 percent of  $4 {\pi} c G / {\kappa}$.   In that 
case the relevant densities are high enough so that absorption 
does increase the effective value of $\kappa$. 
Then the maximum allowed $L/M$ is appreciably 
less than one would expect from the usual formula.  Moreover, 
unlike the classical limit, $\kappa$ is temperature-dependent, 
which may help incite the instabilities.  This line of thought 
originated in connection with LBVs or S Doradus stars (see Section 5 
of Humphreys \& Davidson 1994), but reappears in de Jager's (1998) 
and Lobel's (2001) discussions of yellow hypergiants, albeit 
expressed differently.  Those authors emphasize other considerations 
such as partial ionization and non-radial pulsation, but $L/M$ is 
the essential parameter that makes these stars behave more violently 
than normal supergiants of lower luminosity.\footnote{ 
    \citet{Shav00} has warned that the Eddington limit has been 
    oversimplified in other, geometrical ways, as well, in the past.}  
 
   IRC +10420 is uniquely pertinent in several respects.  It is 
close to the place in the HR diagram where the low-temperature 
edge of de Jager's ``yellow void'' intersects the empirical upper 
luminosity boundary (see Fig.\ 14), and it has a higher luminosity 
and mass loss rate than other known yellow hypergiants. 
In a sense this object vindicates the ``modified Eddington 
limit'' concept:  its wind is dense enough so that 
temperature-dependent Balmer continuum absorption contributes 
substantially to the photospheric opacity, and it is manifestly 
unstable.  In other words, its atmosphere did {\it not\/} become 
tenuous in an asymptotic approach to the classical Eddington limit.  
Is its $L/M$ ratio large enough?  For a  plausible assessment, 
suppose that the effective average $\kappa$ is about 
$1.2 {\kappa}_{sc}$, and that an instability arises when 
$L/M \; {\approx} \; 0.9 \, {\times} \, (4 {\pi} c G / {\kappa}) 
  \; {\approx} \; 33000$;  
then the indicated mass would be about 15 $M_{\odot}$.  The star 
would need to have lost roughly 25 $M_{\odot}$ of its original mass, 
consistent with estimates based on other considerations
(see Papers I and II, and also Nieuwenhuijzen \& de Jager 2000). 

\subsection{The Spectroscopic Connection}

   The structure of the wind should provide a link between
spectroscopy and the star's evolutionary status.  Unfortunately,
the spectral features discussed in Section 3 and by previous 
authors have so far failed to give us a clear geometrical picture.  
Some of them appear almost paradoxical.  Below we suggest a 
phase-change ``rain'' model, but let us approach the problem
in stages.

  First consider IRC +10420's absorption lines:  where are they
formed?  Some are inverse P Cygni components, while others 
resemble features in a quasi-static atmosphere.  Lines formed 
near or below the sonic point in this dense wind should be greatly
broadened by Thomson (electron) scattering, but the observed 
absorption features are fairly narrow.  Three possible 
interpretations are -- 
\\ (1) Maybe the present-day mass loss rate is less than
10$^{-4}$ $M_{\odot}$ yr$^{-1}$, violating the ``proviso'' mentioned
earlier, so the wind is transparent after all.  This recourse 
seems unattractive because the emission lines and rapid
spectral changes suggest an extraordinarily dense wind.
\\ (2) Conceivably the wind flows from only limited regions 
of the star's surface, reminiscent of the solar 
wind with its coronal holes.  If so, then a relatively 
normal atmosphere might exist in the uncovered areas.
This unconventional possibility is motivated by remarks
in subsection 5.4 below.
\\ (3) Or perhaps the absorption lines arise in the wind,
e.g. at radii around 1.5 $R_{star}$ as suggested above
in connection with a simple semi-quantitative model.
We adopt this as the provisional ``best bet'' choice.
\\ -- But then the absorption line velocities present a
difficulty.  If formed in a simple wind, they should have 
negative Doppler shifts relative to the systemic velocity 
of the star, but observations show that this is not the 
case.  The inverse P Cygni absorption components are redshifted 
compared to their related emission, and most other, normal-looking
absorption lines are redshifted by about 15--25 km s$^{-1}$ 
compared to the OH and CO emission (see Section 3 above).  

In principle the observed OH and CO line centers need not coincide 
with the star's systemic velocity as most authors have assumed; 
but even so, the absorption lines discussed in Section 3 are certainly 
not blueshifted by $-20$ to $-60$ km s$^{-1}$ relative to the star as one 
would expect for this wind.   Therefore, as \citet{Oud95}  concluded 
from the inverse P Cygni lines, some sort of ``infall'' seems to dominate 
the absorption-line gas.  He suggested that this is material ejected in
the RSG phase, and which moved out a few hundred AU (or perhaps less)
before falling back.  Dust is expected to form at  
$r \, \sim \, 100$ AU, 10--20 years after ejection; 
so radiation pressure acting on the dust presumably makes infall 
difficult at larger radii.  Velocities must be adjusted within 
fairly narrow limits to enable delayed infall, and why 
a significant fraction of the earlier wind should be falling back on 
the star at the present time is quantitatively unclear.  High-mode 
radial pulsation might provide another explanation for simultaneous 
outflow and infall.  However, we emphasize again that the spectrum is  
formed in a wind rather than a quasi-static atmosphere;  so a radial
pulsation model would need to explain why the local wind falls back
promptly when the stellar surface recedes below it.  Therefore both
qualitative conjectures -- delayed infall and radial-pulsation infall
-- seem unpromising to us.  In the next paragraph we propose a 
different reason why infall may occur in the inner wind.  This idea 
makes use of the special temperature range found there.

\subsection{Outflow and Infall} 

   Based on  quasi-LTE precedent, one expects a phase change 
from ionized to atomic hydrogen to occur near radius 
1.5 $R_{star}$ $\sim$ 1.5 AU, where $T$ falls below 7000 K and
$n_H \; {\sim} \; 10^{11.3}$ cm$^{-3}$.   Inside that zone,
radiation pressure nearly cancels gravity so other effects 
(see, e.g., de Jager 1998) can help radiation to accelerate 
the gas outward.   However, outside the critical radius, opacity 
drastically falls as the electrons recombine; then radiation 
pressure can no longer compete with gravity and the net acceleration 
turns decisively inward.  We doubt that the gas simply coasts out to the 
dust-formation distance from there, since few if any features show 
relative velocities as large as the pertinent escape speed, roughly 
100 km s$^{-1}$.  Instead, {\it conditions in the transition zone 
appear favorable for a phase-change instability which allows infalling 
material to coexist with outflow.\/}  This idea is  speculative but 
physically plausible.  Consider a localized region that has the same 
pressure as its surroundings but slightly lower temperature and higher 
density, and therefore lower ionization.   NLTE heating/cooling 
effects may induce thermal instability,  but the novel 
circumstance here is the {\it opacity decrease\/} in the low-ionization 
blob, which can thus experience a net inward acceleration 
(i.e., gravity dominates) while the surrounding 
lower-density, more ionized, higher-opacity gas is still being 
accelerated outward.   In other words, we propose that 
less-ionized ``droplets'' form in the mostly-ionized gas,
and fall back toward the star until they re-ionize.  A
terrestrial analogy is condensation of rain in an updraft
within a cumulus cloud.  We further suggest that some or most
of the absorption lines form in the inhomogeneous, locally
denser falling material.   This would help explain the observed 
relative redshifts in the IRC +10420 spectrum and perhaps their 
variability in the inverse P Cygni profiles.  NLTE ionization 
may allow continued outward acceleration of the lower-density 
gas flowing outward between the droplets.\footnote{
    The situation described here  pictorially resembles 
    an infall model proposed by \citet{How00} for
    $\tau$ Sco,  but the physical processes are quite different. 
    Their instability results from line-driven acceleration
    in a hot stellar wind, while ours concerns the behavior
    of ionization and average opacity in a critical temperature
    range.  We use the words ``rain'' and ``droplets'' specifically
    to connote a phase change.  However, Fig.\ 1 of Howk et al.\ 
    illustrates our model quite well if the text labels are
    suitably altered.} 
Meanwhile the infalling blobs seem likely to create numerous 
localized shocks wherein temperatures briefly rise to 15000--20000 K.
Magnetic fields may also play a role as we note later. 

   Unfortunately this scenario is extremely difficult to model
quantitatively, and we doubt that any existing computer code 
can provide either a trustworthy demonstration or a disproof. 
The calculations will require NLTE ionization and excitation 
(especially for the $n$ = 2 level of hydrogen), radiative transfer, 
and gas dynamics or even MHD in a conspicuously inhomogeneous 
wind.  A one-dimensional approach cannot simulate the intrinsically 
three-dimensional ``rain.''    All traditional-style atmosphere 
analyses (see Lobel 2001, de Jager et al.\ 2001, 
Nieuwenhuijzen \& de Jager 2000, and refs.\ cited therein) 
therefore appear unsuitable for this case.  Lobel's quantitative 
result, that the averaged adiabatic index ${\Gamma}_1$ can be low 
enough to cause instability in cool hypergiants, is relevant 
and suggestive but it offers little guidance for the detailed
structure of a brisk dense wind like that of IRC +10420.  

  We offer some remarks concerning size scales in the 
``rain'' zone.  As noted above, that region is located around 
$R \; {\sim} \; 1.5 R_{star}$   and the thickness of the ionization 
transition zone is expected to be of the order of $0.15 R$.  
The two most obvious intensive dynamical parameters, 
the speed of sound $w$ and net gravitational acceleration 
$g_{eff} \; {\sim} \; 0.5 g$, suggest a likely order-of-magnitude 
size for an infalling blob or ``droplet'':
$$    
w^{2}/g_{eff} \; {\sim} \; 5 {\times} 10^{11} \; {\rm cm} \; {\sim} \; 0.02 R.
$$
This dimensional-analysis result is essentially the same as the 
scale height that a static atmosphere would have at the same radius.  
The proposed size is optically thick in the core of the H$\alpha$ 
line (see below).  The internal timescale for a blob (i.e., size/$w$) 
would be a few days and its infall-and-reionization lifetime is 
of order 10--50 days.  

   One might expect the double-peaked H$\alpha$, H$\beta$, and Ca~II 
emission to provide clues to the structure of the wind.  Our STIS observations 
show that at least the H$\alpha$ profile appears more or less the same 
from all directions, and therefore does not represent a circumstellar 
disc or ring or polar outflow as previously thought (Section 4 
above).   \citet{deJ97a} found that an ensemble of turbulent 
shocks can produce a double-peaked H$\alpha$ profile, but their example
is much narrower and weaker than that observed in IRC +10420.  Most 
of the H$\alpha$ from this object emerges at blueshifts and redshifts 
around ${\pm}$60 km s$^{-1}$, well outside the thermal and turbulent 
width.  The blue peak $F_{\lambda}$ is at least 8 times the continuum 
level (Fig.\ 4), indicating an origin either in 
gas hotter than 20000 K, or else at radii of the order of $4 R_{star}$, 
or a mixture of both high temperature and large overall extent.  
Thus it is not obvious whether the H$\alpha$ emission originates 
in the localized inflow/outflow shocks mentioned above (reminiscent 
of the de Jager et al.\ case) or, alternatively, in an extended 
outflow zone at large radii. This ambiguity is enhanced by the fact 
that H$\alpha$ is rather {\it weak\/} relative to the high mass 
loss rate assumed here.  The total  H$\alpha$ luminosity\footnote{ 
    Assumptions for this estimate:  The measured equivalent width
    is 44 {\AA}, and the intrinsic continuum energy distribution
    resembles a 9000 K black body with a luminosity of 
    $10^{5.7} \; L_{\odot}$.} 
is roughly 2000 $L_{\odot}$, which in a nebular context would 
require a volume emission measure of the order of 
$10^{61.2}$ cm$^{-3}$.  But this is tiny compared to the 
V.E.M.\ $>$ 10$^{64}$ cm$^{-3}$ that the wind would have 
if it were fully ionized;  evidently the average ionization 
fraction must be quite low in the outer wind, $r \; > \; 1.5 R_{star}$. 
Thus, both obvious possibilities -- emission from many 
small hot shocked zones in the innermost wind where infall
occurs, or else nebula-like emission from widespread parts of the 
outer wind -- are much harder to assess than H$\alpha$
production in a hotter, fully ionized wind.  The very broad 
line wings (Fig.\ 4) give another clue:  
at least a substantial fraction of the H$\alpha$ 
emission comes from regions with noticeable scattering by
free electrons, i.e., with column densities between 10$^{23}$ 
and 10$^{24}$ cm$^{-2}$.  As noted earlier this is probably 
not true for the absorption lines.   In summary,  the H$\alpha$ 
problem remains murky and some quantitative details 
seem to contradict each other!

   Why do the H$\alpha$, H$\beta$, and Ca~II lines have double
peaks?  \citet{deJ97a} do not clearly explain this result for 
their turbulent shock example.  Our suggestion for IRC +10420 is 
that critical regions have large velocity gradients and relatively 
small thickness in the radial direction, allowing line photons
to escape from their creation zones along the outward and 
inward directions but not sideways.  This idea seems to 
fit well with the conjectured local shocks between infall
and outward flow in a ``rain'' scenario.  Its applicability
in the outer regions, however, is less clear.  

\subsection{The possible importance of stellar activity} 

   Finally, we note that magnetic fields and stellar activity (in 
the same sense as solar activity) may play a role in this 
story.  Convection and turbulence can occur more easily and more 
violently near the upper luminosity boundary in the HR diagram, 
conceivably even in stars hotter than 10000 K if they are close enough
to the Eddington limit.  The case is strongest in the red supergiant 
stage.  Star-spots, magnetic fields, and associated outflow are 
known to occur in RSG's such as $\alpha$ Ori and VX Sgr (see, e.g., 
\citet{Sch75,Gil96,Klu97,Uit98,Lobd00,Lobd01,Tut97,Chap86,Tri98}). 
It would not be very surprising if such activity has large-scale 
dynamical importance in the most extreme cases.  The very luminous 
cool supergiant VY CMa has ejected material in a few directional 
looplike structures, qualitatively reminiscent of solar prominences 
but immensely larger and more massive \citep{Smith01}.  Non-radial
pulsation might provide an alternative origin for these structures,
but this idea seems less appealing since they are narrow, few in
number, and remarkably looplike in appearance.   If dynamically
significant turbulent/magnetic activity does occur, then one obvious 
question is:  How long does it  persist during post-RSG evolution?  
The same factors that make a yellow hypergiant surface unstable 
lend themselves to the generation of turbulence and magnetic fields. 

Images of the IRC +10420 ejecta show numerous arcs, knots, and jetlike
structures (Fig.\ 1 and Paper II) which indicate localized ejection
events in seemingly random directions, analogous to VY CMa although 
the two morphologies are not identical.  As mentioned in Section 4,
the velocity trend across one of these arcs suggests a flat geometry  
--  which would be unexpected if it was created by non-radial pulsations, 
but reasonable if it was shaped by a magnetic field.  Located at 
radii from 0.5$\arcsec$ to 2$\arcsec$ (2500 to 10000 AU), these features 
were probably ejected between 400 and 1000 years ago.  Although it 
is difficult to know how far the star had evolved at those times, 
at least the innermost features may have been formed after 
the star had ceased to be an extremely cool RSG (see Section 6 below).  
\citet{Ned92} have reported surprisingly strong magnetic fields of 
the order of 1 mG in the OH maser regions at $r \, \sim$ 7000 AU. 
A conventional extrapolation inward  
($B^2 \; \propto \; {\rho}^{4/3}$ or $B \; \propto \; r^{-2}$)
would give $B \; \sim$ many kG at the current stellar surface,  
impossibly strong because the magnetic energy density would then 
exceed all other local energy densities by a huge factor.  A field 
proportional to $r^{-1}$, on the other hand, would give a surface 
magnetic energy density comparable to the thermal energy density 
(Section 5.1 above).  

Given the critical temperature range and general instability 
\citep{Lob01}, we should not be surprised if dynamically significant
convective/magnetic activity occurs above the surface of IRC +10420 in 
its currently observed state.   The thermal energy of the inner wind 
described in Section 5.1 exceeds that of the solar photosphere by a 
factor of order 10$^6$ while $v_{esc}^{2}$ is smaller by a factor of 
100, and the region is dynamically less stable than the solar 
photosphere and chromosphere.  Major surface activity therefore seems 
possible or even likely.  This idea does not conflict with the 
rough global isotropy of the outflow (Section 4.1 above), 
since the hypothetical sporadic events may be numerous and random 
in direction, consistent with observed structures in the 
ejecta.  Observations have been too sparse to detect possible 
rapid variations in spectral details.   In summary, the ejecta of 
VY CMa and IRC +10420 provide credible evidence that turbulent/magnetic
activity may help to cause or at least shape ejection events in 
the most extreme cool supergiants.  Quantitative models, however,  
would be even more difficult to construct than for the ``rain'' 
phenomenon proposed above.  

   Active magnetic fields, and also shocks like those proposed 
in Section 5.3, can accelerate charged particles to non-thermal 
energies.   Solar flares produce MeV particle energies and indirectly 
produce neutron fluxes \citep{For86,Ram86}.  Therefore nuclear 
processing in the surface layers might conceivably affect the 
abundances of rare elements in spectra of extreme supergiants 
such as IRC +10420 and VY CMa. This remark is intended merely as 
a speculation that needs quantitative assessment in the future.

\section{Final Remarks -- Crossing the Yellow Void}

From our perspective the yellow hypergiants, hidden by their dense
winds, may appear to be relatively stalled on the upper HR diagram in
the temperature range 6000 -- 8000$\arcdeg$K.  Nieuwenhuijzen and de Jager
(1997, 1998) said that  these objects  were ``bouncing'' at the cool edge of
the ``yellow void'', but that  expression  refers only to the wind or 
superficial outer layers of the star. If the intermediate type
hypergiants are post--red supergiants on blueward evolutionary tracks,
as we believe for IRC +10420, then  their  interiors or cores will
continue to evolve, unconcerned with their  external appearance. 
So how can a yellow hypergiant cross the yellow void?

With these apparent temperatures and their very high luminosities, the
yellow hypergiants are in the dynamically unstable region in the HR diagram
described in the previous section and experience high mass loss rates and 
discrete ejection episodes with even higher mass loss accompanied by 
spectroscopic and photometric variations.  The historical and 
recent shell episodes observed in $\rho$ Cas are examples 
\citep{Bid57,Beard61,Lob02}. The very luminous F-type star Var A 
in M33, which now shows TiO bands \citep{RMH87}, is very likely 
an example of a similar but more extreme and much longer 
shell event. Figure 14 shows an HR diagram with the yellow void 
and the  positions of IRC +10420 and Var A plus $\rho$ Cas and HR 8752 
from \citet{Isr99} and the  apparent temperature shifts corresponding 
to their recent spectroscopic variations.  Unlike IRC+10420, HST/WFPC2 
images of $\rho$ Cas and the very similar hypergiant HR 8752 show no 
evidence for circumstellar material \citep{Sch02}. Thus HR 8752 and 
$\rho$ Cas may have only recently encountered this unstable region in 
their blueward evolution. Furthermore, the normal mass loss rates for 
HR 8752 and $\rho$ Cas are at least 10 times less than  
those reported for IRC +10420.  

Our HST images of IRC+10420 provide us with a snapshot of its mass loss 
history and show evidence for several mass loss episodes. The outermost 
reflection arcs at $\approx$ 5$\arcsec$ were ejected about 3000 years ago,
undoubtedly when the star was a red supergiant as this is comparable to, 
or greater than, the  time expected for a massive star star to evolve 
back to warmer temperatures from the RSG stage \citep{Sch92,Sch93} 
The complex structures closer to the star correspond to more recent 
high mass loss, possibly discrete, asymmetric events. In Paper II we 
showed that IRC +10420 experienced  a high mass episode during the past 
600 years shedding about 1 $M_{\odot}$ with a 
mass loss rate $\sim$ 10$^{-3}$ $M_{\odot}$ yr$^{-1}$. IRC +10420 has 
undoubtedly been in this highly unstable region of the HR diagram for
several years.

The time scales for post red supergiant evolution, especially at these 
high luminosities, are very uncertain, ranging from several hundred to 
a few thousand  years \citep{Sch92,Sch93}.  \citet{Sto01} have suggested 
a model in which these stars {\it appear\/} to make frequent rapid 
excursions across the HR diagram due to dynamical instabilities 
as red supergiants. Whether post red supergiant evolution 
is smooth and continuous or sporadic,  
the yellow hypergiants very likely remain in this unstable state 
until, due to interior evolution, the underlying surface, defined as the 
sonic point beneath the opaque wind, becomes sufficiently hot 
($\ge$ 12000$\arcdeg$K) to shed its dense false-photosphere 
or opaque wind. The star would next be seen on the blue side 
of the void with a temperature of 12000--20000$\arcdeg$K,
possibly as a post-RSG LBV like the "less luminous" type in Figure 14
\citep{HD94} and with a much reduced mass loss rate.  In principle, 
if the wind is opaque enough, the underlying stellar surface 
{\it might\/} already be above 12000 K;  we do not propose that this
is the case, but mention the possibility as a reminder of the 
disconnection between spectrum and stellar radius when a dense wind exists.

In addition to its vigorous mass loss earlier, IRC +10420 has shown 
some dramatic changes during the past century and even during just 
the last 30 years. Its historical light curve shows that it brightened
by a magnitude in the 50 years prior to 1970. This might be due to a shift
in bolometric correction to warmer temperatures after a shell episode
lasting decades similar to Var A, although most authors (Paper I, Oudmaijer
et al 1986) have interpreted it as thinning of the circumstellar dust
along the line of sight. \citet{Blo99} show that a two-component
dust shell model based on their near--infrared speckle--interferometry 
``can be interpreted as evidence of a termination of an enhanced mass 
loss episode 60 -- 90 years ago'', which interestingly corresponds 
to the onset of the visual brightening. Thus, it appears that a high 
mass loss episode ended only recently  and that its mass loss rate, while
still high, has declined during  the past few decades.  The apparent
temperature of an opaque wind tends to increase as the mass loss rate 
decreases \citep{KD87}.  With the 
apparent warming of its dense wind, the recent appearance of strong 
hydrogen emission, and a possible decline in its mass loss rate, we 
suggest that IRC +10420 may be {\it in transit across the semi-forbidden 
region of the yellow void.} However, we do not know whether it has just 
begun or is near the end of its passage.  The latter is the more 
interesting possibility, and continued observations of this remarkable 
star in the next few decades may reveal its future evolution.

\acknowledgments
Support was provided by NASA through grant number GO-7304 from the
Space Telescope Science Institute, which is operated by the
Association of Universities for Research in Astronomy, Inc., under
NASA contract NAS 5-26555. We are especially grateful to Kazunori
Ishibashi for processing the HST/STIS spectra of IRC+10420 with the
STIS IDT software. We also thank Rene Oudmaijer for his comments on the
inverse P Cygni profiles, Kerstin Weis for useful discussions
on the spectra of unstable massive stars, and Robert Lysak  for conversations
about the Sun and solar activity.

\newpage

\begin{figure}
\plotone{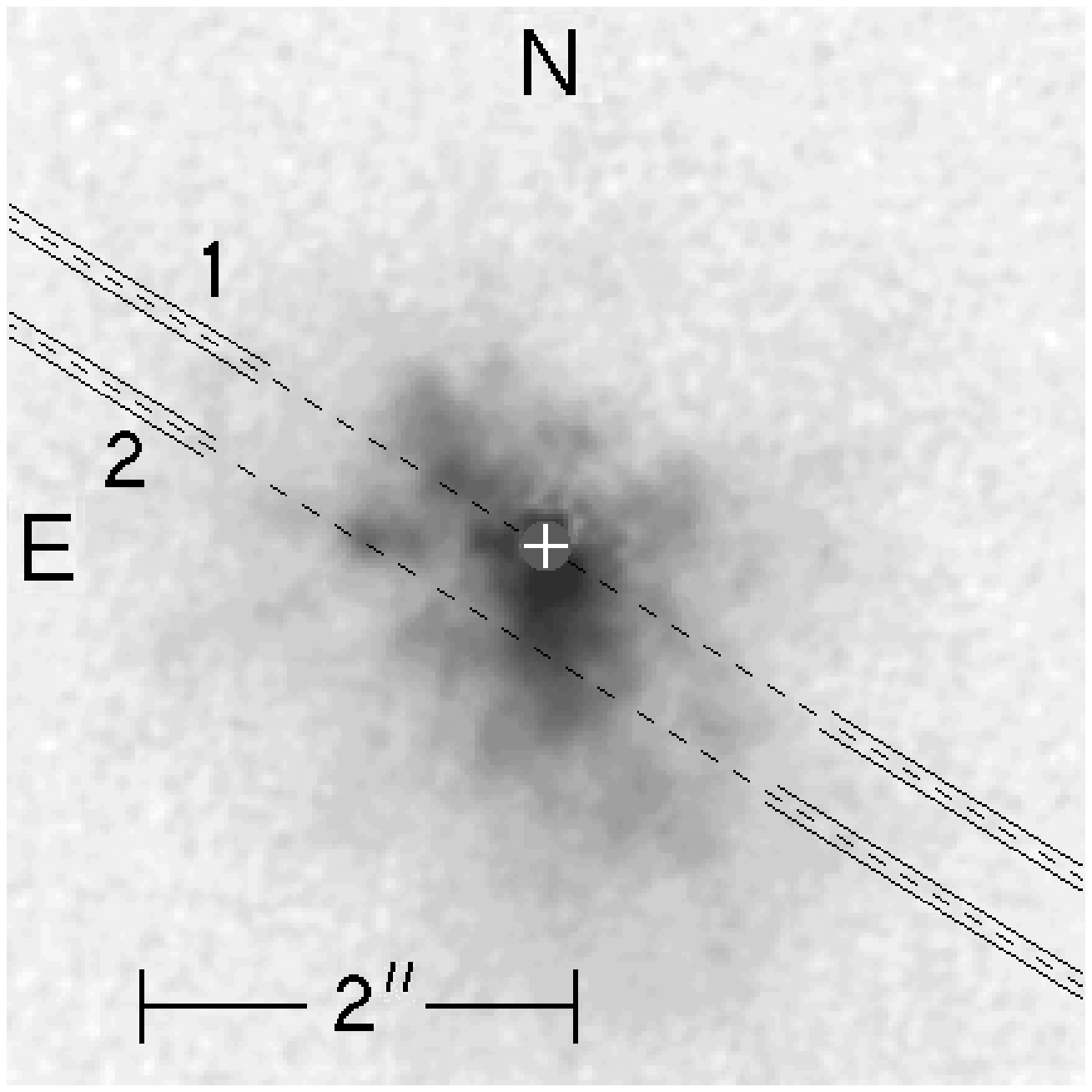}
\caption{Positions of the two HST/STIS slits on the IRC +10420 WFPC2 image 
  from Paper II.}
\end{figure}

\begin{figure}
\plotone{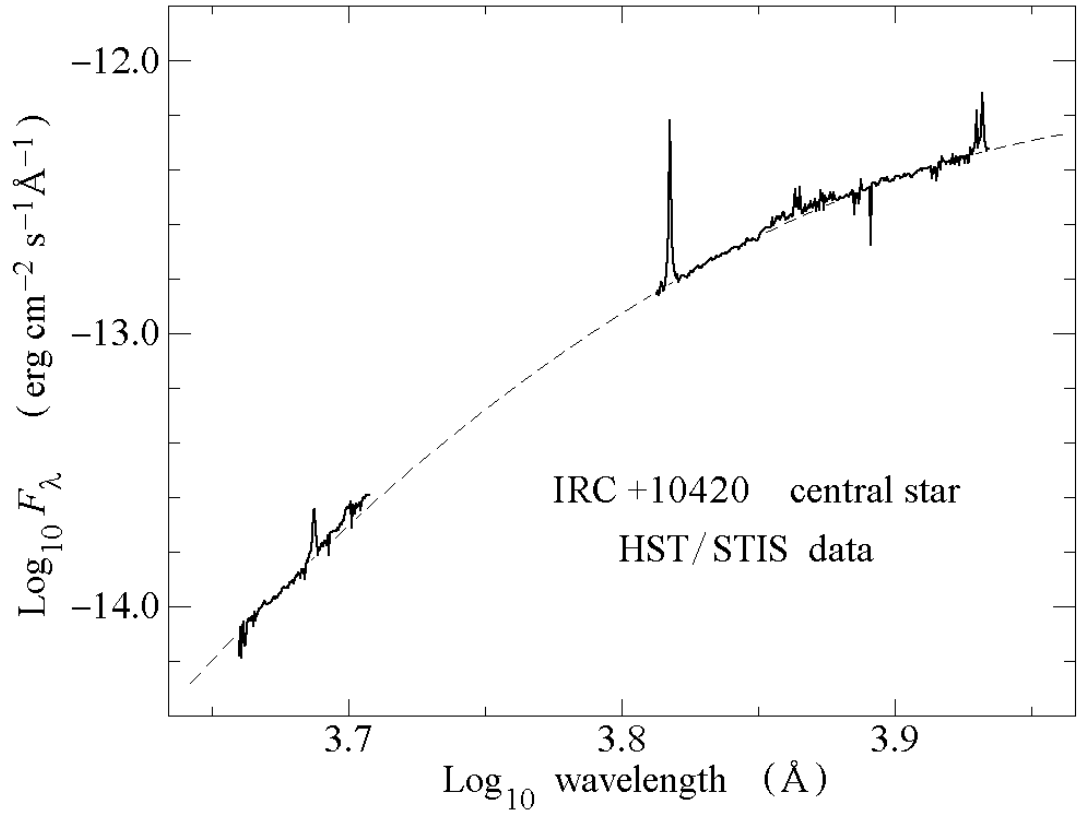}
\caption{The flux calibrated spectrum of the central star with the parabolic 
  fit to the continuum.}
\end{figure}

\begin{figure}
\plotone{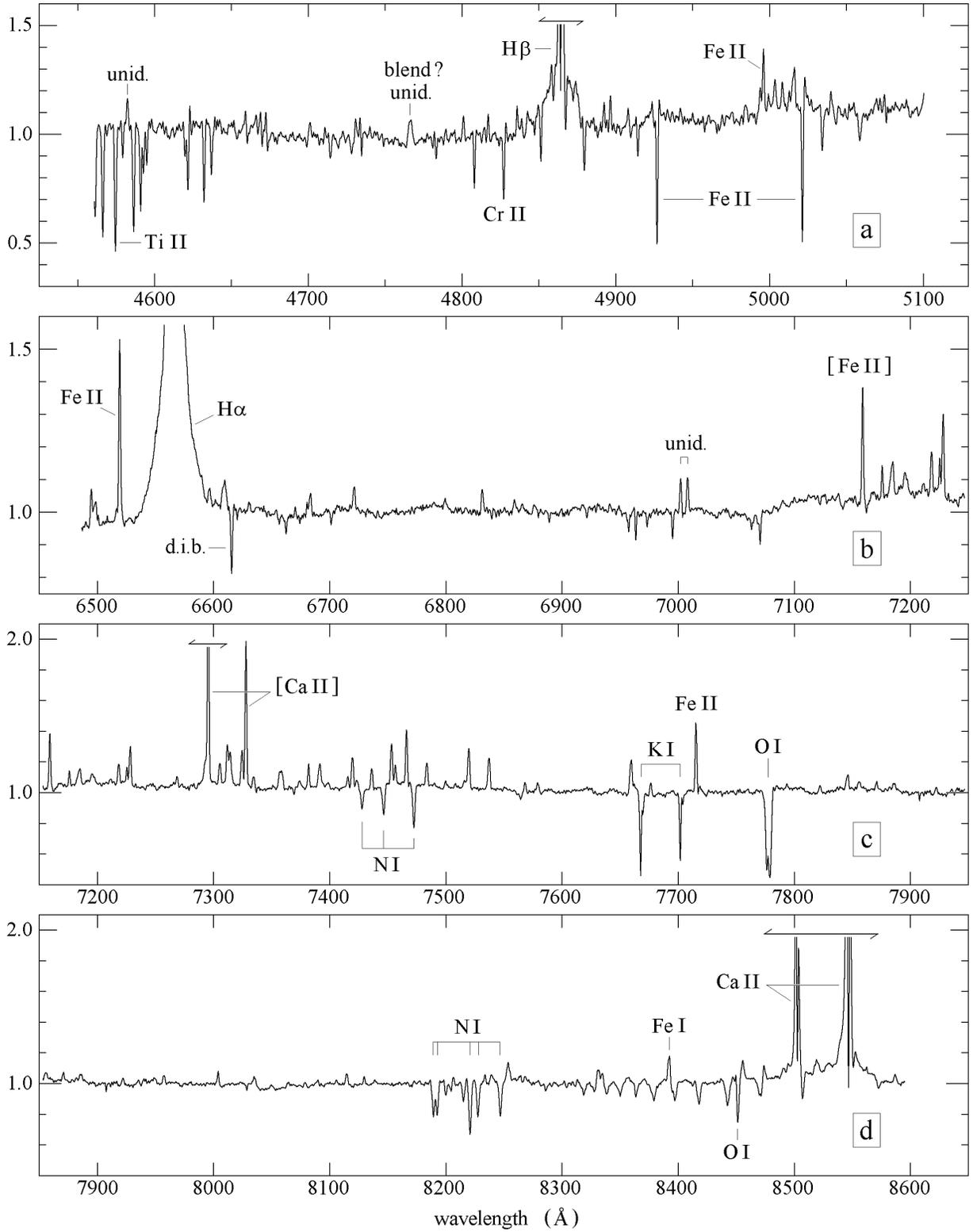}
\caption{The normalized blue and red spectra of the central star with some 
  of the stronger lines identified.}
\end{figure}

\begin{figure}
\plotone{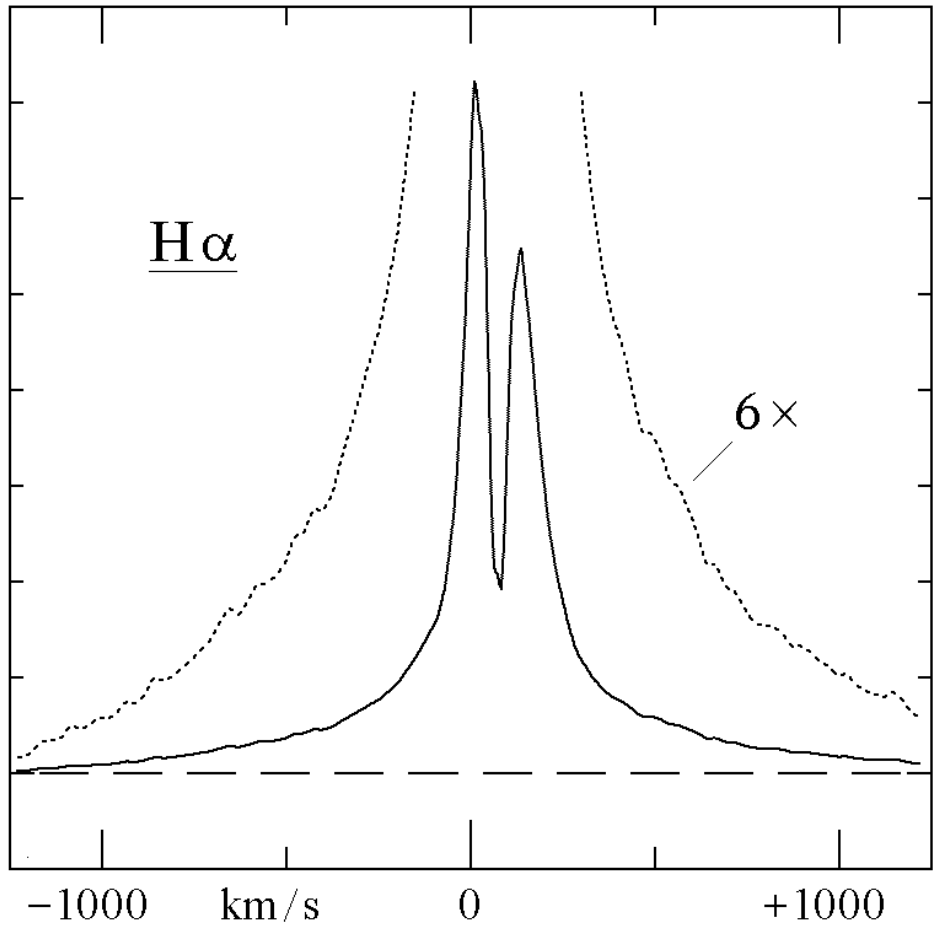}
\caption{The profile of the H$\alpha$ emission line showing the broad electron 
  scattering wings projected against a 6$\times$ enlargement of the inner profile.}
\end{figure}

\begin{figure}
\plotone{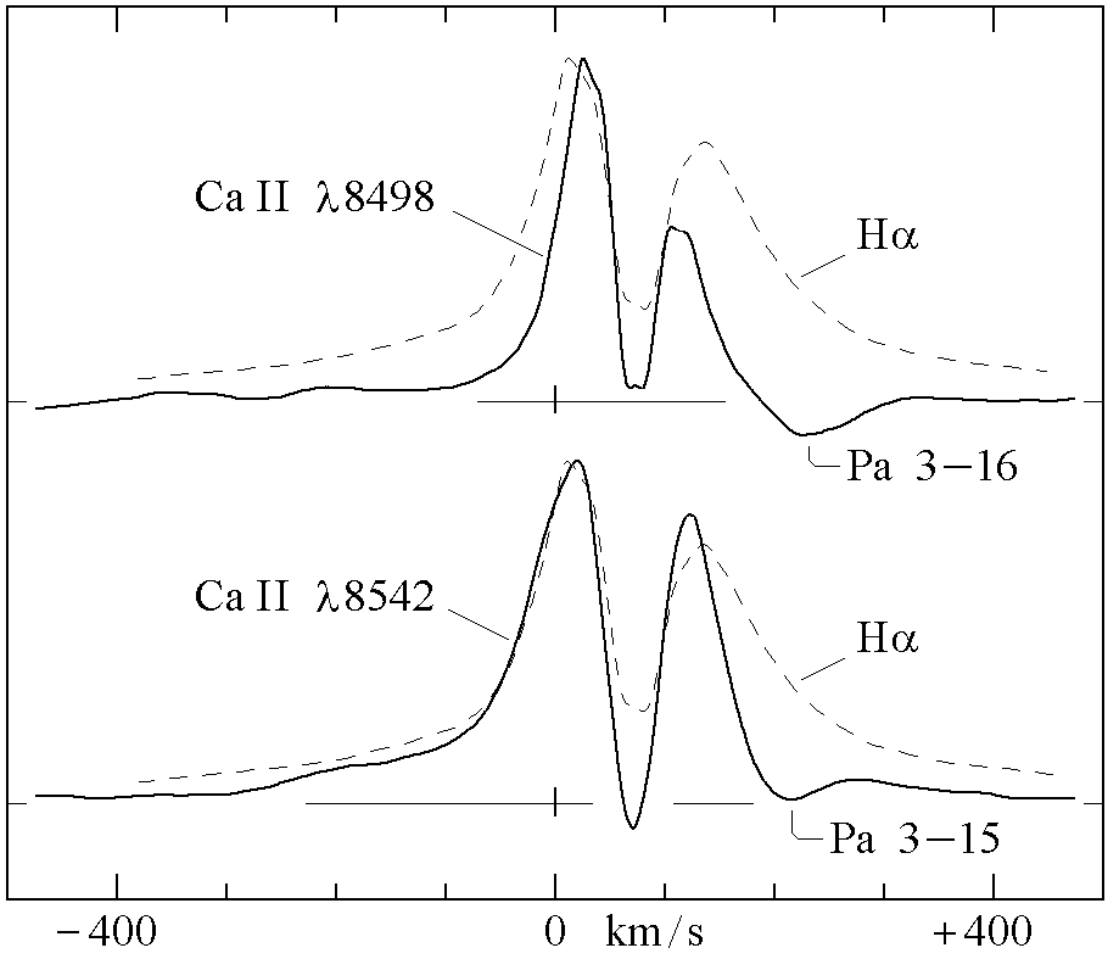}
\caption{The split emission line profiles of the two Ca II lines at 
  $\lambda$8498 and $\lambda$8542 compared with the H$\alpha$ profile.}
\end{figure}

\begin{figure}
\plotone{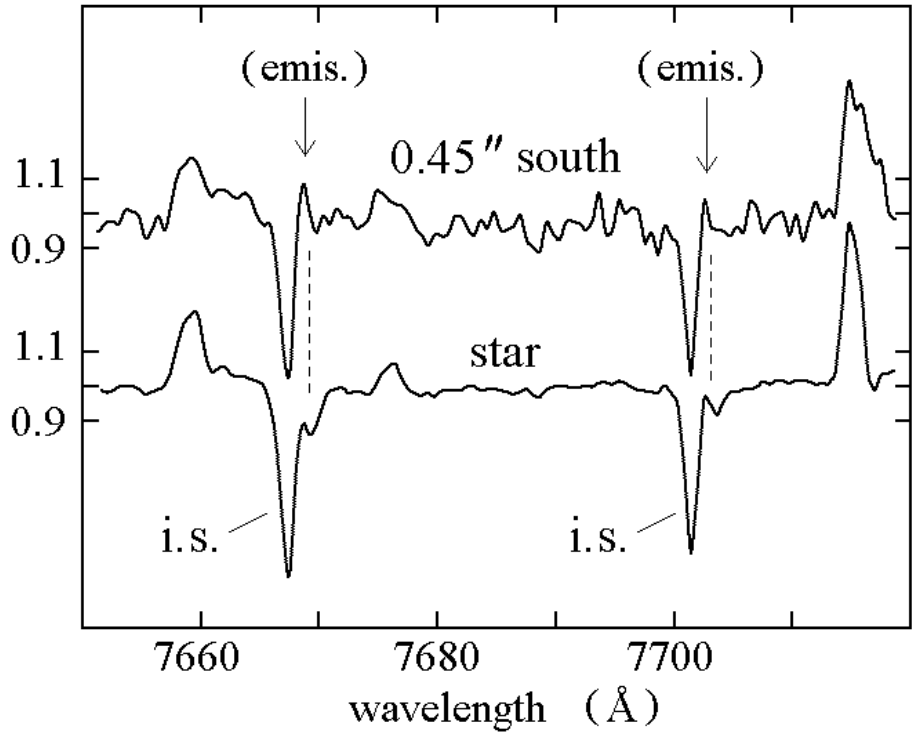}
\caption{The wavelength region showing the K I emission lines. The strong 
  interstellar K~I absorption line and the K I absorption in the star are also 
  identified.  The lower panel shows the spectrum on the central star and the 
  upper panel shows the same wavelength region in the ejecta at offset position 
  $\underline{a}$ (Figure 12). The K~I emission is clearly present above the 
  continuum level in the ejecta.}
\end{figure}

\begin{figure}
\plotone{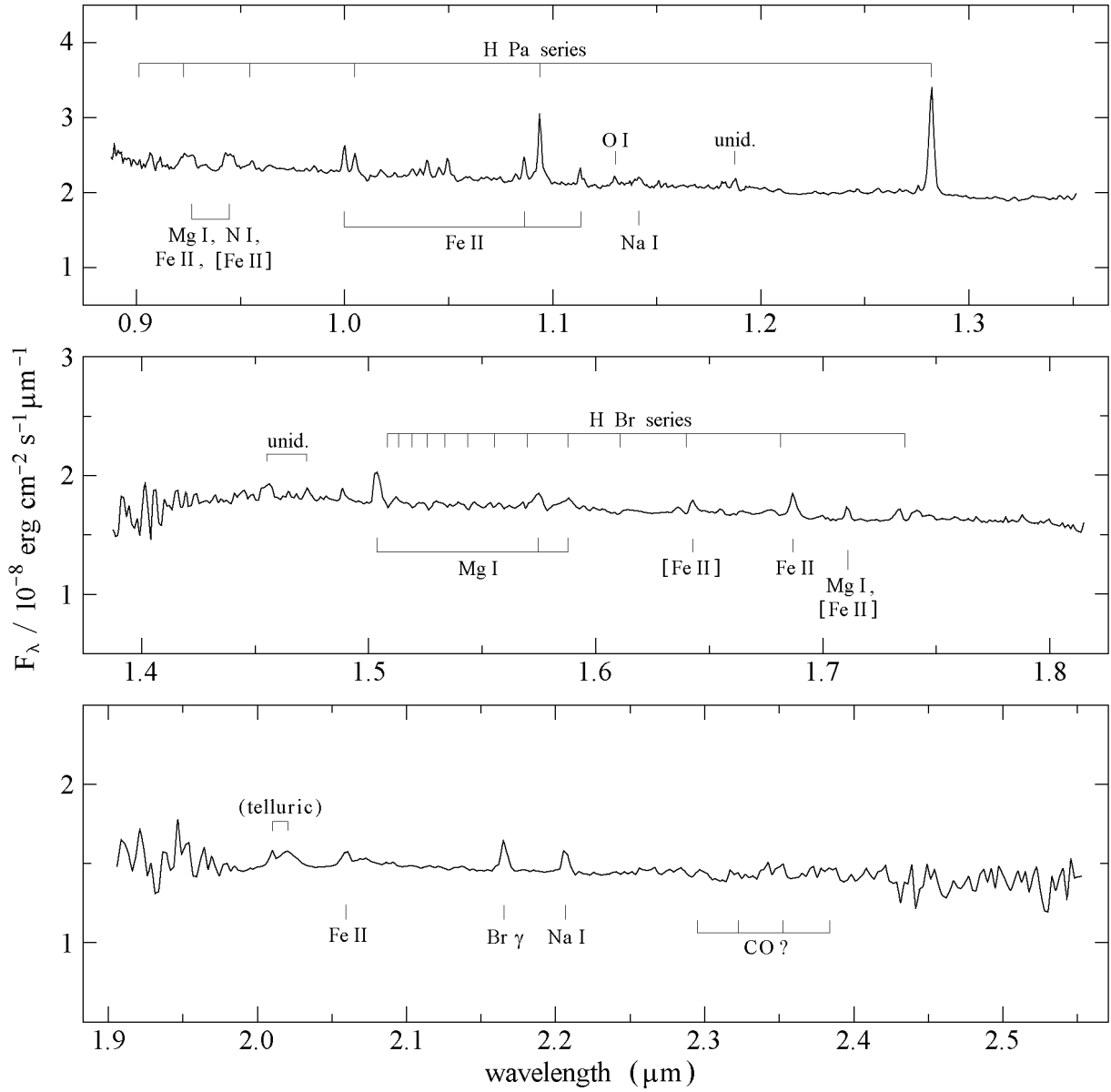}
\caption{Low-resolution near-IR spectrum of  IRC+10420 observed with CRSP on the 
  KPNO 2.1m telescope. The signal to noise is worst at the edges of each bandpass,
  especially in the K-band, due to poor atmospheric transmission and filter 
  cut-offs. In this figure ``unidentified'' means the resolution is inadequate 
  to choose among several possibilities.}
\end{figure}

\begin{figure}
\plotone{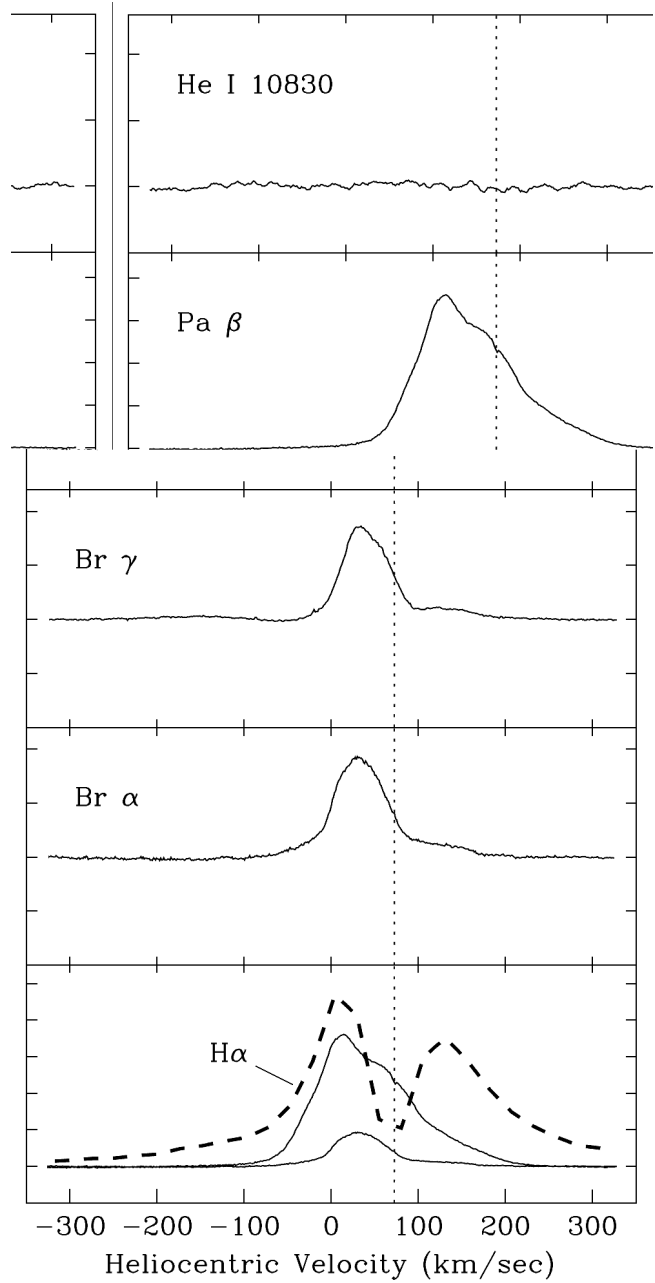}
\caption{Line profiles of near-IR emission lines observed with PHOENIX on the 
  KPNO 4m telescope.  The presumed heliocentric radial velocity of +73 km s$^{-1}$ 
  is shown by the dotted vertical line in each panel.  The bottom panel compares  
  near-IR hydrogen line profiles observed with PHOENIX to the H$\alpha$ profile
  of the central star observed with HST/STIS (dashed line) } 
\end{figure}

\begin{figure}
\plotone{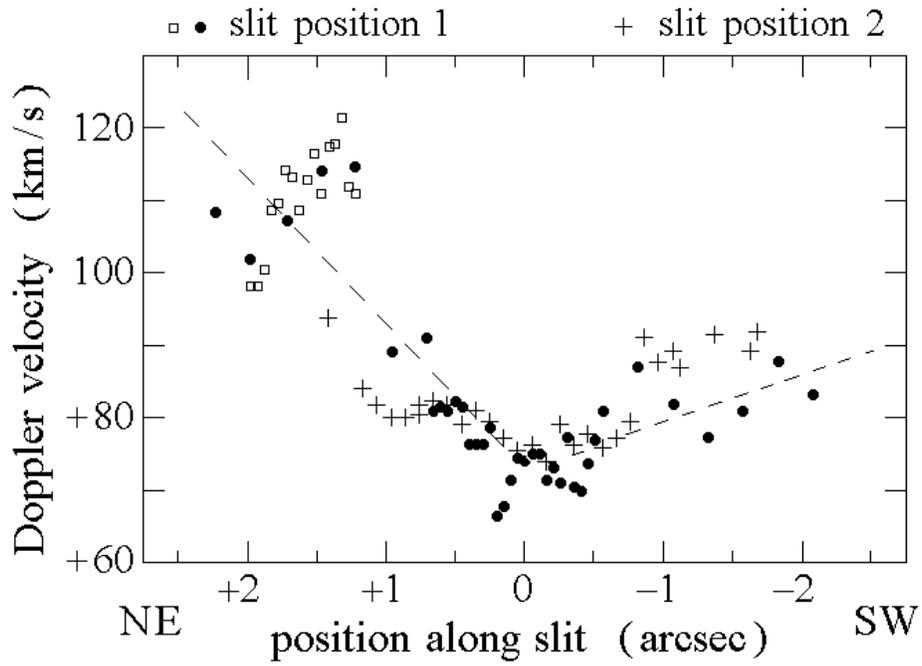}
\caption{The measured velocities along the two slit positions. The measurements 
  along slit positions 1 and 2 are shown as filled circles and crosses 
  respectively. The open squares are the measurements across the arc at slit 
  position 1.}
\end{figure}

\begin{figure}
\plotone{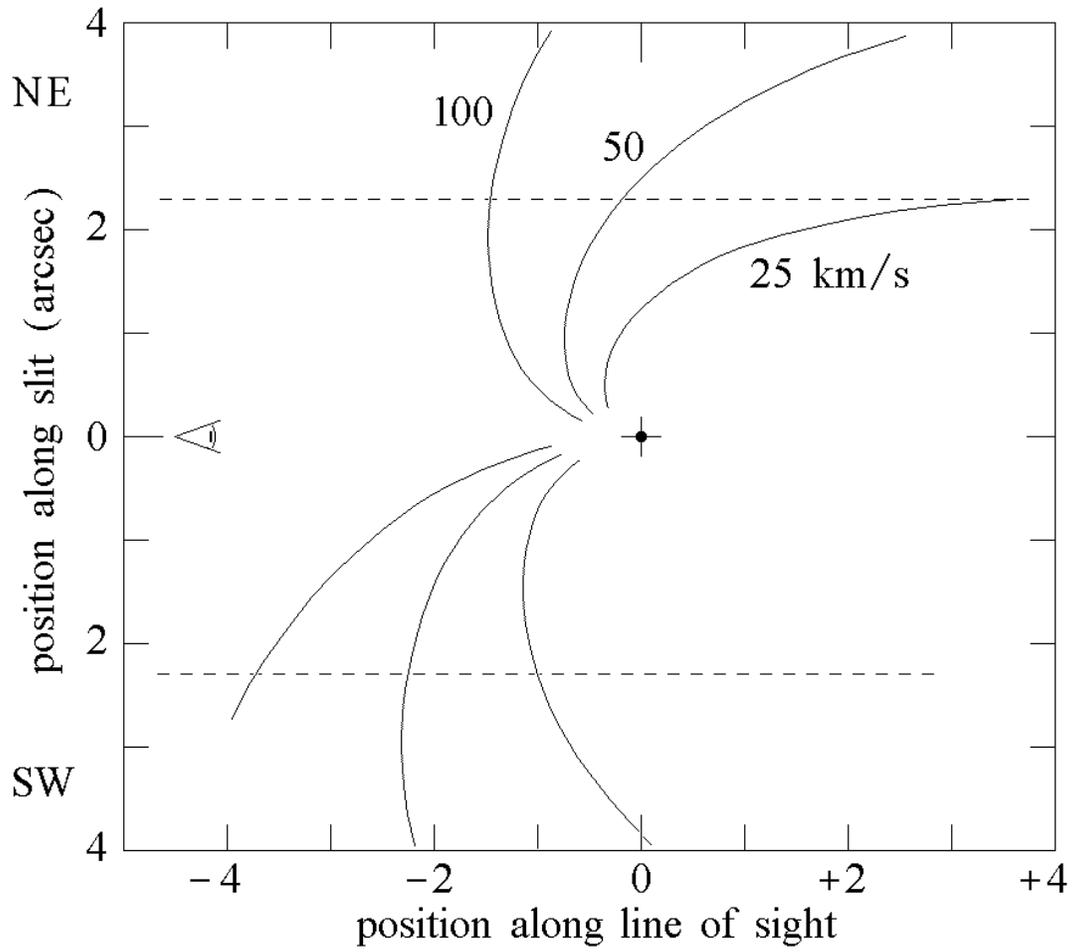}
\caption{Each set of two curves shows the spatial loci corresponding to the 
  velocity trends in Figure 9 for three different expansion velocities. Our 
  viewpoint is to the left.  The curves are not valid within 0$\farcs$5 of
  the center, partly because of scatter in the data there (cf.\ Fig.\ 9).}
\end{figure}

\begin{figure}
\plotone{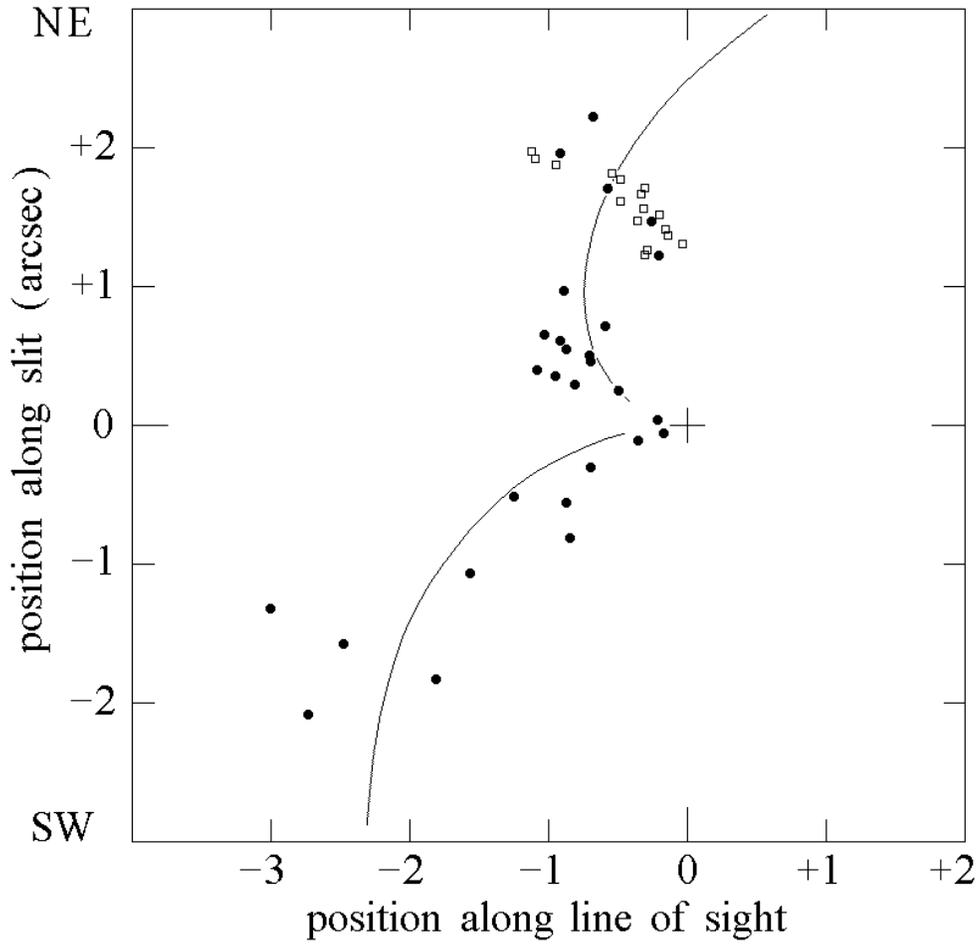}
\caption{Spatial distribution of the data points in Figure 9 for 
  V$_{r}$ = 50 km s$^{-1}$. The symbols are the same as in Figure 9.}
\end{figure}

\begin{figure}
\plotone{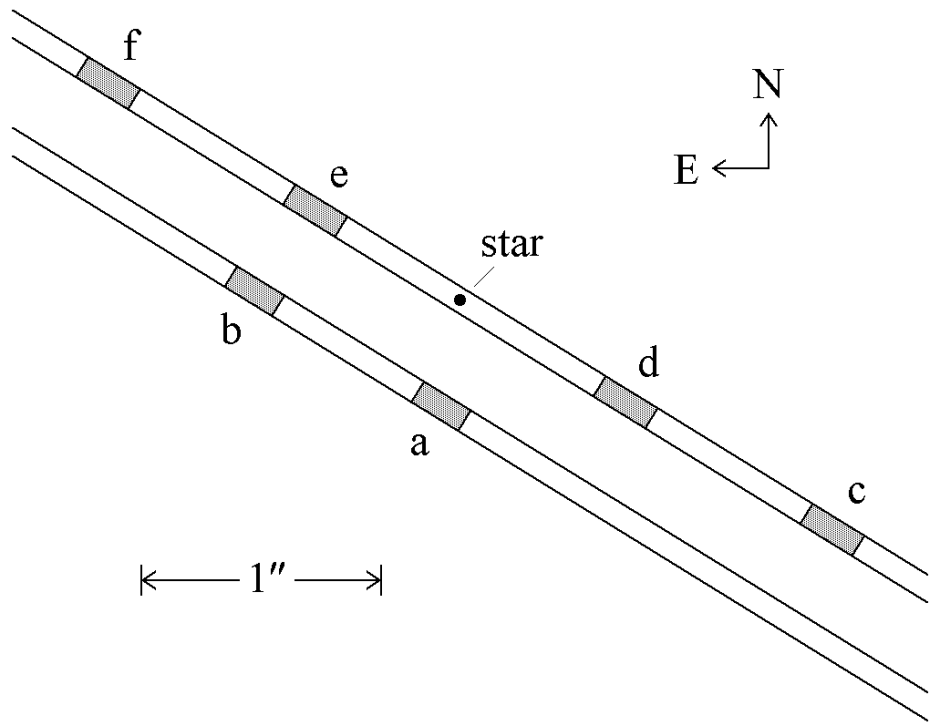}
\caption{Map of the 6 positions used to sample the H$\alpha$ profile.}
\end{figure}

\begin{figure}
\plotone{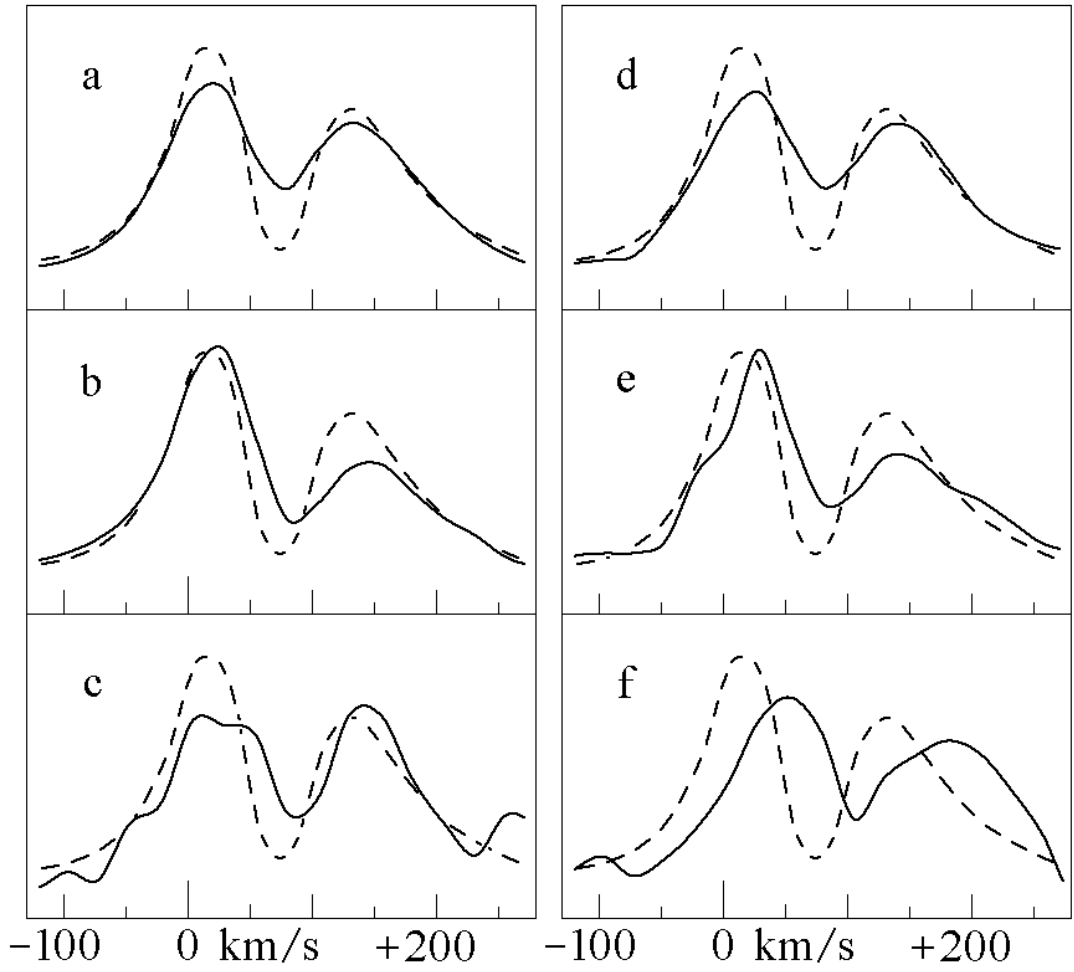}
\caption{The H$\alpha$ profiles at the 6 different positions in the ejecta. 
  For comparison,  the H$\alpha$ profile on the central star is shown with a 
dashed line.}
\end{figure}

\begin{figure}
\plotone{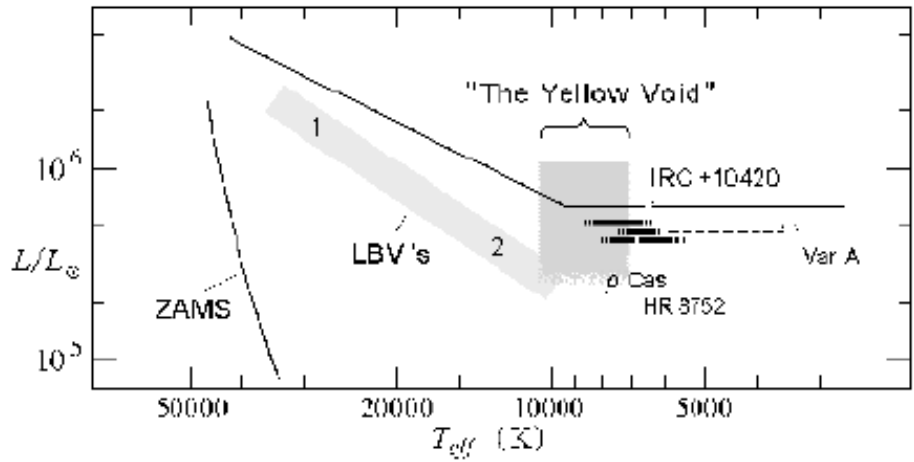}
\caption{A schematic HR diagram showing the position of the yellow void with
  IRC +10420, Var A in M33, $\rho$ Cas, and HR 8752 and their range in apparent
  temperature corresponding to their spectral variations (see text). The shell
  episodes in Var A and $\rho$ Cas are shown as dashed lines. Locations of the
  classical (1)  and less luminous (2) LBV's in their quiescent state are shown as
  a band, also known as the the S Doradus instability strip.  The solid line is
  the empirical upper luminosity boundary.}
\end{figure}

\clearpage 
 
\begin{deluxetable}{ccccc}    
\footnotesize
\tablecaption{Journal of HST/STIS  Observations (September 10, 1999). \label{tbl-1}}
\tablewidth{0pt}
\tablehead{
\colhead{Grating} & \colhead{Slit Position} & \colhead{$\lambda_{c}$} & \colhead
{Exp. Time(sec)} & \colhead{${\lambda_{c}}/{\Delta}{\lambda}$}  
}
\startdata
G750M &1 &6768 &8, 30, 432 &6150\\ 
   &2 &     &   432     &    \\ 

    &1 &7283  &   432    &6620\\  
    &2 &      &          &    \\  

   &1 &7795  &   288    &7090\\ 
   &2 &      &          &     \\  

   &1 &8311  &   324    &7560\\ 
   &2 &      &          &    \\  

G430M &1 &4706  & 60, 1440 &8400\\  

   &1 &4961  &   600    &8860\\ 
   &2 &      &171, 2592 &    \\   
\enddata
\end{deluxetable}

\clearpage 

\begin{deluxetable}{ccc}   
\footnotesize 
\tablewidth{285.06732pt}
\tablecaption{Near-Infrared Spectroscopy . \label{tbl-2}}
\tablehead{
  \colhead{Filter} & \colhead{$\lambda$($\micron$)\tablenotemark{a}} & 
  \colhead{${\lambda_{c}}/{\Delta}{\lambda}$\tablenotemark{b}} 
  }
\startdata
\sidehead{CRSP Observations, 13 June 2000} 
I & 0.99, 1.09 & 600 \\
J & 1.22 & 490\\
H & 1.54, 1.66 & 700\\
K & 2.22 & 440\\

\sidehead{PHOENIX Observations, 17 -- 19 June 2000}
He I J9232 & 1.083 & 108000\\
Pa$\beta$ J7799 & 1.282 & 94000\\
Br$\gamma$ K4578 & 2.166 & 108000\\
Br$\alpha$ L2462 & 4.053 & 126000\\

\enddata
\scriptsize 
\tablenotetext{a}{If two central wavelengths are given (i.e. I and H), two grating tilts were used to cover the bandpass}
\tablenotetext{b}{The resolution corresponds to the 2--pixel detector resolution; for PHOENIX is actually somewhat worse, $\sim$ 75000} 
\end{deluxetable}

\clearpage

\begin{deluxetable}{cccccc}    
\footnotesize
\tablecolumns{6}
\tablewidth{0pt}
\tablecaption{Apparent Magnitudes and Colors of IRC +10420. \label{tbl-3}}
\tablehead{
\colhead{} & \multicolumn{2}{c}{Central Star} & \colhead{} & \multicolumn{2}{c}{Star + halo}\\ 
\cline{2-3} \cline{5-6} \\
\colhead{Mag or Color} & \colhead{STIS\tablenotemark{a} 1999.7} & \colhead{WFPC\tablenotemark{b} 1996.3} &   \colhead{} & \colhead{WFPC\tablenotemark{b} 1996.3} & \colhead{Ground-based\tablenotemark{c} 1991} 
}
\startdata
B  & 14.76\tablenotemark{d} & 14.6: & & 13.57: & 13.63 \\
V  & 12.18\tablenotemark{d} & 11.9: & & 11.06: & 10.93 \\
R$_{CK}$ & 10.45 & \nodata & & \nodata & 9.56\\
I$_{CK}$ &  8.90 & \nodata & & \nodata & 7.95\\
\\
B--V  & +2.58 &  +2.7: & & +2.5: & +2.70 \\
V--R  & +1.73 &  \nodata & & \nodata & +1.37\\
V--I  & +3.28 &  \nodata & & \nodata & +2.98\\
\enddata
\scriptsize 
\tablenotetext{a}{This paper} 
\tablenotetext{b}{Fluxes reported in Paper II}
\tablenotetext{c}{Paper I, V = 11.03 in 1994 \citet{Oud96}}
\tablenotetext{d}{Extrapolated or interpolated, see text and Fig.\ 2} 
\end{deluxetable}

\clearpage

\begin{deluxetable}{cccc}   
\footnotesize 
\tablewidth{0pt}
\tablecaption{Heliocentric Velocities\tablenotemark{a} of the Emission Lines with Split Profiles. \label{tbl-4}}
\tablehead{
\colhead{Line} & \colhead{Blue Em.} & \colhead{Absorption} & \colhead{Red Em.}
}
\startdata
H$\alpha$   & 14.1  & 72.3  & 134.2\\

H$\beta$    & 23.4  & 71.3  & 127.7\\

Ca II $\lambda_{vac}$8500  & 22.3  & 71.6  & 113.1\\

Ca II $\lambda_{vac}$8544  & 9.2  & 69.2  & 122.8\\
\enddata
\scriptsize
\tablenotetext{a}{To convert to LSR velocities add 17.5 km s$^{-1}$} 
\end{deluxetable} 

\clearpage

\begin{deluxetable}{cccc}   
\footnotesize
\tablewidth{0pt}
\tablecaption{Mean Heliocentric Velocities of the Absorption and Emission Lines. 
   \label{tbl-5}}
\tablehead{
\colhead{Lines} & \colhead{Mean Velocity}  & \colhead{Number of Lines} & \colhead{Remark}
}
\startdata
\sidehead{Absorption Lines:}
N I &  74.1 $\pm$ 4.4 & 10\\

O I &  76.1 $\pm$ 7.8 & 3\\

Paschen lines & 75.8 $\pm$ 3.9 & 11\\

K I (stellar) & 92.4 $\pm$ 5.3 &  2\\  
K I (IS)      & 12.7 $\pm$ 1.7 &  4\\

Fe II,Ti II,Cr II, etc. &  86.2 $\pm$ 8.5 & 20\\  

\sidehead{Emission Lines:}
[Ca II] &  68.5 $\pm$ 1.2 & 2\\ 

Permitted   &  67.2 $\pm$ 8.1 & 28 & Fe II,Sc II,Y II, etc.\\ 

Forbidden   & 64.4 $\pm$ 7.9 & 8 & Fe II,V II,Cr II, etc.\\  

Neutral(Fe I) & 68.7 $\pm$ 10.2 & 3\\  

Neutral(K I, slit 2) & 61.0 $\pm$ 3.2 & 2\\ 
\sidehead{Inverse P Cygni Profiles:}

Emission    &  40 $\pm$ 7  & 13\\

Absorption & 107 $\pm$ 7  & 13\\
\enddata
\end{deluxetable} 

\clearpage

\begin{deluxetable}{ccccc}   
\footnotesize
\tablewidth{448.11697pt}
\tablecaption{Additional Line Identifications in the Spectrum of the Central Star. 
\label{tbl-6}}
\tablehead{
  \colhead{Element \& Mult.} & \colhead{$\lambda$$_{air}$}  & 
  \colhead{W$_{\lambda}$} & \colhead{Hel.\ Vel.\tablenotemark{a}} & 
  \colhead{Notes}
  }
\startdata
\sidehead{Absorption Lines:}
V II  29 & 4844.31 & 0.09 & +84.8 & 1. \\
                                                     
\sidehead{Emission Lines:}
Fe II (26) & 4580.07 & -0.21 & +58.8 & 2.\\

Sc II (13)\tablenotemark{b} & 4698.28 & -0.10 & +68.3 & 1.\\
 
Ti II (48)\tablenotemark{c} & 4763.84 & -0.20 & +61.5 & 2.\\
 
Fe II (30) & 4825.71 & -0.04 & +74.9 & 1.\\
 
[Fe II] (4F) & 4889.63 & -0.10 & +64.8 & blended with [Fe II] 3F, 3.\\
 
[Fe II] (3F) & 4889.70 &   & +60.5 & blended with [Fe II] 4F, 3.\\
 
[Fe II] (20F) & 4905.4 & -0.06 & +61.7 & 2.\\
 
[Ti II] (23F) & 4925.84 & -0.08 & +47.5 & 1. \\
 
Y II (20)\tablenotemark{d} & 4982.13 & -0.12 & +71.4  & blended with [Ti II] 23F, 2.\\
 
[Ti II] (23F)\tablenotemark{d} & 4982.73 &   & +53.6 & blended with Y II 20, 2.\\
 
[Fe II] (20F) & 5005.5 & -0.17 & +73.4 & 2.\\
 
Ti II (113) & 5010.20 & -0.09 & +61.7 & 1.\\
 
Ti II (113)\tablenotemark{e} & 5013.38 & -0.33 & +44.9 & 2.\\
 
[Fe II] (20F) & 5020.2 & -0.17 & +68.1 & 1.\\
 
Fe II (36) & 5036.92 & -0.07 & +82.2 & 2.\\
 
[Fe II] (19F)\tablenotemark{f} & 5072.40 & -0.05 & +38.9 & 2.\\

[Ti II] (8)  & 6592.93 & -0.07 & +78.3 & 1.\\ 

[Fe II] (14)\tablenotemark{g} & 7171.98 & -0.11  & +66.8 & 2.\\ 
 
[V II] (4F)\tablenotemark{c} & 7353.77 & -0.42 & +70.9 & 3. \\
 
[V II] (4F) & 7387.5 & -0.41 & +63.0 & 1.\\

[Fe II] (14)\tablenotemark{g} & 7388.16 & -0.36  & +60.0 & 2.\\ 

Mn II (4) & 7415.80 & -0.42 & +62.8 & 3.\\
 
[Fe II] (47) & 7432.23 & -0.24 & +64.8 & 3.\\                    

[Fe II] (14)\tablenotemark{g} & 7452.50 & -0.20 & +64.7 & 2.\\ 

Fe II & 7866.53 & -0.14 & +65.2 & 1.\\

Y II (32) & 7881.90 & -0.16 & +57.5 & 2.\\

Fe II & 7917.80 & -0.06 & +71.6 & 3.\\

\enddata
\end{deluxetable} 

\clearpage

\begin{deluxetable}{ccccc}    
\footnotesize
\tablenum{6}
\tablewidth{448.11697pt} 
\tablecaption{Table 6 -- continued}
\tablehead{
  \colhead{Element \& Mult.} & \colhead{${\lambda}_{air}$}  & 
  \colhead{W$_{\lambda}$} &
  \colhead{Hel.\ Vel.\tablenotemark{a}} & \colhead{Notes}    
  }
\startdata

Fe II & 8031.32 & -0.13 & +60.5 & 3.\\

[Ti II] (6F) & 8060.16 & -0.06 & +60.8 & 1. \\
 
[Ti II] (6F) & 8085.2 & -0.06 & +51.8 & 1.\\

Mn II & 8110.39 & -0.15 & +69.5 & 2.\\

Fe II & 8451.00 & -0.28 & +74.5 & 1.\\
\enddata
\scriptsize
\tablenotetext{a}{add 17.5 km s$^{-1}$ for LSR velocity}
\tablenotetext{b}{possible blend}
\tablenotetext{c}{very broad, blend}
\tablenotetext{d}{very broad, blend; three peaks at lab rest wavelengths of 4981.02\AA, 4981.91\AA, 4982.81\AA}
\tablenotetext{e}{very broad, blend; two  peaks at lab rest wavelengths of 5012.45\AA and 5013.54\AA}
\tablenotetext{f}{Possible inverse P Cygni profile with neighboring absorption line which would have a velocity of +118.3 km s$^{-1}$. If not, then the absorption line is unidentified at lab rest wavelength of 5072.96\AA}
\tablenotetext{g}{Identified by Oudmaijer (1998) with [Fe I]}  
\tablecomments{1. not apparent in Oudmaijer's (1998) spectrum, 2. visible in Oudmaijer's (1998) spectrum, but not in the line list, 3. listed as unidentified in Oudmaijer (1998)}
\end{deluxetable}          

\clearpage

\begin{deluxetable}{ccccc}   
\footnotesize 
\tablewidth{337.08438pt}
\tablecaption{Unidentified Lines in the STIS Spectra of the Central Star. \label{tbl-7}}
\tablehead{ 
\colhead{Measured $\lambda$} & \colhead{$\lambda$$_{vac}$} & \colhead{$\lambda$$_{air}$}
& \colhead{W$_{\lambda}$} & \colhead{Notes}
}
\startdata 
\sidehead{Absorption Lines:}
4698.79  & 4697.45 & 4696.13 & 0.04 & 1.\\

4713.70  & 4712.35 & 4711.03 & 0.06 & 2.\\

4714.48  & 4713.13 & 4711.81 & 0.09 & 2.\\

4748.77  & 4747.41 & 4746.08 & 0.05 & 1. \\

4787.91  & 4786.54 & 4785.20 & 0.04 & 1. \\

4872.22  & 4870.83 & 4869.46 & 0.07 & 1.\\

4909.69 &  4908.28 & 4906.91 & 0.21 & 1.\\

6888.96 &  6887.00 & 6885.05 & 0.04 & 1.\\

6957.41 &  6955.42 & 6953.47 & 0.08 & 2.\\

6963.53 &  6961.53 & 6959.58 & 0.11 & 2. \\ 

6973.26 &  6971.26 & 6969.31 & 0.08 & 2.\\

\sidehead{Emission Lines:}

6609.09\tablenotemark{a} &  6607.62 & 6605.77 & -0.27 & 3.\\

6831.29 &  6829.78 & 6827.87 & -0.13 & 3. \\ 

6858.89 &  6857.37 & 6855.45 & -0.10 & 3. \\

7002.11 &  7000.56 & 6998.60 & -0.17 & 3.\\

7008.10 &  7006.55 & 7004.58 & -0.20 & 3.\\

7195.36 &  7193.76 & 7191.75 & -0.49 & 1.\\

7268.36 &  7266.74 & 7264.71 & -0.13 & 1.\\

7381.59 &  7379.95 & 7377.89 & -0.24 & 3.\\

7567.85 &  7566.17 & 7564.05 & -0.14 & 3.\\

7578.91 &  7577.23 & 7575.11 & -0.15 & 3. \\

7676.03 &  7674.32 & 7672.17 & -0.14 & 2. \\

7821.92 &  7820.18 & 7817.99 & -0.10 & 3.\\

7855.63 &  7853.89 & 7851.69 & -0.21 & 1.\\

7957.16 &  7955.40 & 7953.17 & -0.11 & 3.\\

8078.55 &  8076.75 & 8074.49 & -0.03 & 1. \\

8105.67 &  8103.87 & 8101.60 & -0.05 & 1. \\

8473.70 &  8471.82 & 8469.45 & -0.14 & 1. \\ 

8529.09\tablenotemark{b} &  8527.20 & 8524.81 & -0.32 & 1. \\ 

8552.36 &  8550.46 & 8548.07 & -0.35 & 1.\\ 

8562.14 &  8560.24 & 8557.84 & -0.05 & 1. \\ 
\enddata 
\scriptsize
\tablenotetext{a}{A broad, probably blended, emission line with peaks at 
  6604.67 {\AA} and 6606.45 {\AA} in the laboratory rest frame. A possible 
  line identification is Sc~II 6604.60 {\AA} (19). Oudmaijer (1998) shows 
  this feature resolved into four emission components.}
\tablenotetext{b}{Broad line}  
\tablecomments{ 1.\ not apparent in Oudmaijer's (1998) spectrum, 2.\ visible 
  in Oudmaijer's (1998) spectrum but not in the line list, 3.\  listed as 
  unidentified in Oudmaijer (1998)}
\end{deluxetable}

\clearpage

\begin{deluxetable}{ccccc}   
\footnotesize
\tablewidth{0pt}
\tablecaption{Absorption Line Differences Between the 
Star and the Ejecta.  \label{tbl-8}}
\tablehead{
\colhead{Element \& Mult.} & \colhead{$\lambda$$_{air}$} & \colhead{W$_{\lambda}$ star} & \colhead{W$_{\lambda}$ ejecta} & \colhead{Remark}
}
\startdata 
Fe II (42) & 5018.4 & 0.57 & 0.18 & \\

Fe II (42) & 4923.9 & 0.60 & 0.03 & \\

& & & &  \\

Cr II (30) & 4824.1 & 0.37 & 0.11 & \\

Cr II (30) & 4848.2 & 0.25 & 0.15 & \\

Cr II (30) & 4864.3 & 0.32 & 0.18 & \\

Cr II (30) & 4876.4 & 0.33 & 0.24 & blend \\

& & & &  \\

unidentified & & 0.11 & 0.015 & $\lambda$6963.5, see Table 7\\
\enddata  
\end{deluxetable}

\clearpage

\begin{deluxetable}{ccc}  
\footnotesize
\tablewidth{337.08438pt}
\tablecaption{Emission Line Identifications\tablenotemark{a} in the Near--Infrared Spectra\tablenotemark{b}. \label{tbl-9}}
\tablehead{
\colhead{Element \& Mult.} & \colhead{$\lambda$$_{air}$ in microns} &\colhead{Comments}
}
\startdata 
Mg I (27)  &  0.9256  &   blended with [Fe II] \\

[Fe II] (13)  & 0.9267  &  blended with Mg I \\

[Fe II] (13) & 0.9399 &  blended with Mg I  \\   

Mg I (38) &    0.9415 &  blended with [Fe II] \\  

[Fe II] (13) & 0.9471 &  blended with Mg I \\ 

Fe II  & 0.9997 & \\  

Fe II & 1.0174 &\\

[S II] (3) & 1.0286  &\\

[N I] (3) & 1.0398, 1.0407    &  blend\\

[Ni II] (11) & 1.0459  & doubtful id.\\

Fe II & 1.0501  &\\

Mg I (37) & 1.0811 &  blend\\ 

Fe II &  1.0863, 1.0872  & blend \\  

Fe II &  1.1125  &\\

O I (7) & 1.1287  & blend\\

Na I (3) & 1.1381, 1.1403   & blend \\

Mg I (6)  & 1.1829 & \\

[Fe II] & 1.2567  & \\ 

Na I (21)  & 1.2679   & \\

[Fe II] & 1.2703  & \\   

[Fe II] & 1.3205  & \\ 

[Fe II] & 1.3278  & \\  

Ca II: & 1.4555  & blend, uncertain id\\

Mg I & 1.5025, 1.5040, 1.5048  & blend\\

Mg I & 1.5136 & blend\\

Mg I & 1.5741, 1.5749, 1.5765  & blend\\

Mg I & 1.5880, 1.5886  & blend\\

Mg I & 1.6365  &\\

[Fe II] & 1.6435  & \\

[Fe II] & 1.6637 &\\

[Fe II] & 1.6766  &\\ 

Fe II & 1.6873  &\\

Mg I & 1.7109  & blend with [Fe II]\\  

Fe II & 1.7338  &\\

[Fe II] & 1.7449  &\\ 

Fe II & 2.0600  &\\

Na I & 2.2056, 2.2077, 2.2084 & blend\\
\enddata
\scriptsize
\tablenotetext{a}{We used \cite{Col93} available from the ADC and the Atomic Line List at http://www.pa.uky.edu/~peter/atomic/}
\tablenotetext{b}{Not including the hydrogen lines}
\end{deluxetable} 

\end{document}